\documentclass[
    aps,
    prb,
    reprint,
    superscriptaddress,
    amsmath,
    amssymb
]{revtex4-2}

\usepackage{comment}
\usepackage{graphicx}
\usepackage{dcolumn}
\usepackage{bm}
\usepackage{braket}
\usepackage{hyperref}
\usepackage{xcolor}

\begin{document}

\title{Scaling Dimensions and Two-Point Amplitudes from Operator Covariance Matrices}

\author{You-Chiuan Chen}
\email{ymc5544@psu.edu}
\affiliation{
Department of Physics, The Pennsylvania State University,
University Park, Pennsylvania 16802, USA
}

\author{Ribhu K. Kaul}
\affiliation{
Department of Physics, The Pennsylvania State University,
University Park, Pennsylvania 16802, USA
}
\affiliation{
Center for Theory of Emergent Quantum Matter, The Pennsylvania State University,
University Park, Pennsylvania 16802, USA
}

\date{\today}

\begin{abstract}
The long distance behavior of two-point functions of critical lattice statistical mechanics models depend on detailed information of the underlying quantum field theory. Yet these universal properties can be difficult to extract because generic lattice operators couple to the infinite hierarchy of continuum scaling operators with unknown coefficients. Recently in  Physical Review E
{\bf 114}, 024146 (2026), one approach to disentangle some of this information was proposed using the eigen-decomposition of the operator covariance matrix constructed from cross correlations of a finite set of $N_\ell$ lattice operators. In this work we analyze this method in detail.  We provide an explicit proof that under the assumption of scale invariance the leading decay of the eigenvalues are controlled by the smallest $N_\ell$ scaling dimensions, even after taking irrelevant perturbations into account. Based on this understanding, we propose a new method by which the eigenvectors of the covariance matrix can be used to construct the amplitudes of the continuum two-point correlation functions. This allows for a complete determination of the structure of the continuum two-point functions from the lattice covariance matrix. We demonstrate our ideas both on synthetic data sets (drawn from the scaling hypothesis) and numerically exact data for the two dimensional Ising model.  
\end{abstract}

\maketitle

\section{Introduction}

A cornerstone discovery of theoretical physics in the last half century is that the long distance physics of microscopic statistical mechanics lattice models at their critical points are described by continuum massless Euclidean quantum field theories. The link between these two descriptions is made in theory through the renormalization group (RG)~\cite{Fisher1998}, though in practice only a limited amount of the full structure of the continuum theory can be extracted successfully from the study of lattice models.

Explicitly, at a critical point, the long-distance correlations of a lattice model are described by an infinite hierarchy of continuum scaling operators $\{\phi_1,\phi_2,\cdots\}$ with scaling dimensions $\{\Delta_1,\Delta_2,\cdots\}$ (arranged in ascending order)\cite{Cardy1996,Wegner1976}. For scalar operators, scale invariance gives their two-point functions the form
\begin{equation}
    \langle\phi_a^{\dagger}(\mathbf{0})\phi_b(\mathbf{r})\rangle
    =\frac{k_{ab}}{r^{\Delta_a+\Delta_b}}.
    \label{eq:continuum_correlator}
\end{equation}
where $\dagger$ denotes the operator adjoint. Notice the scaling dimensions $\Delta_a$ and the two-point amplitudes $k_{ab}$ constitute two pieces of universal data: the former determine the power-law decay of the correlations, while the latter characterize the correlations between the scaling operators. We normalize our operators in the standard way, so that $k_{aa}=1$. The continuum theory has more structure in higher point  functions, but in this work we focus on the two-point functions. The central question we are interested in here is: How does one extract the universal data $\Delta_a$ and $k_{ab}$ from lattice correlation functions data?

Following Ref.~\cite{Sandvik2026}, but generalizing the formulation to complex-valued operators, we consider $N_\ell$ lattice operators $O_i(\mathbf{x})$ and define their connected covariance matrix by
\begin{equation}
    C_{ij}(\mathbf{x}_1,\mathbf{x}_2) = \langle O^{\dagger}_i(\mathbf{x}_1)O_j(\mathbf{x}_2)\rangle - \langle O^{\dagger}_i(\mathbf{x}_1)\rangle \langle O_j(\mathbf{x}_2)\rangle,
    \label{eq:covariance_definition}
\end{equation}
where $1 \leq i,j \leq N_\ell$. Note that for classical operators that are complex-valued functions, $O^{\dagger}_i(\mathbf{x})$ reduces to $O^*_i(\mathbf{x})$, where $^*$ denotes complex conjugation. Because of translation symmetry on the lattice, the covariance matrix depends only on the separation $\mathbf{r}=\mathbf{x}_2-\mathbf{x}_1$. Therefore,
\begin{equation}
    C_{ij}(\mathbf{r}) = C_{ij}(\mathbf{x}_1,\mathbf{x}_2) = \langle O^{\dagger}_i(\mathbf{0})O_j(\mathbf{r})\rangle - \langle O^{\dagger}_i \rangle \langle O_j \rangle,
    \label{eq:covariance_separation}
\end{equation}
where the local average $\langle O_i(\mathbf{x}) \rangle = \langle O_i \rangle$ is spatially invariant. Translation invariance and complex conjugation imply $C(-\mathbf{r})=C(\mathbf{r})^{\dagger}$. Here we assume inversion symmetry, $C(-\mathbf{r})=C(\mathbf{r})$, so that the covariance matrix is Hermitian.

In Ref.~\cite{Sandvik2026}, Sandvik proposed diagonalizing the symmetric operator  matrix at each separation, using a unitary matrix $U(\mathbf{r})$ such that, $ D(\mathbf{r}) = U(\mathbf{r})^{\dagger}C(\mathbf{r})U(\mathbf{r})$. Here $D(\mathbf{r})$ is diagonal, and we denote its $i$th entry, which is an eigenvalue of $C(\mathbf{r})$, by $\lambda_a(\mathbf{r})=[D(\mathbf{r})]_{aa}$. While the study of the eigen decomposition of such matrices has a long history in the lattice field theory community for extracting particle masses~\cite{Luscher1990}, it appears that Ref.~\cite{Sandvik2026} was the first to consider it in detail for critical phenomena. In this recent work, it was proposed and demonstrated numerically that the leading asymptotic behavior of the eigenvalues is \cite{Sandvik2026}
\begin{equation}
    \lambda_a(r)\sim r^{-2\Delta_a}.
    \label{eq:sandvik_scaling}
\end{equation}
where $\Delta_a$ are the smallest scaling dimensions of scaling operators in the same symmetry sector as the chosen lattice operators. We emphasize that while the original work clearly confirms the result in Eq.~\ref{eq:sandvik_scaling} by checking it with Monte Carlo data for various models, a proof or a complete argument for the validity of this result was not provided. 

Our goal in this manuscript is to provide an explicit proof of Eq.~\ref{eq:sandvik_scaling}, use our proof to go beyond~\cite{Sandvik2026}  by studying the eigenvectors of the operator covariance matrix, and test all our results in detail numerically. The manuscript is organized as follows: Our first objective is to give an explicit proof for why diagonalization of the covariance matrix uncovers the hierarchy of scaling dimensions,   Eq.~\ref{eq:sandvik_scaling}, including finite degeneracies and corrections from irrelevant perturbations. Our proof relies on scale invariance at the critical point and is given explicitly in a step-by-step manner in Sec.~\ref{sec:setup}. Our second objective is to show that the eigenvectors of the covariance matrix, $U({\bf r})$ can be used to extract the universal two-point amplitudes $(k_{ab})$ of the resolved scaling operators, a method to do this is proposed in Sec.~\ref{sec:covariance_eigenvectors_theory}. Taken together we show that the entire universal structure of the two point function, Eq.~\ref{eq:continuum_correlator} can be extracted from the operator covariance eigen-decomposition. In the final two sections  we provide explicit and detailed test of these ideas, first on synthetic data in Sec.~\ref{sec:numerics} and then on the 2D Ising model in Sec.~\ref{sec:ising_application}. Detailed proofs of the linear algebra results and further demonstrations of our results are relegated to appendices.

\section{Covariance Eigenvalue and Scaling dimensions, $\Delta_a$}
\label{sec:setup}

In this section we explicitly prove the eigenvalue scaling, Eq.~\eqref{eq:sandvik_scaling} proposed by Sandvik in a few steps. Our proof only assumes the form of correlation functions of scaling operators Eq.~\eqref{eq:continuum_correlator} and the expansion of lattice fields in the scaling operators.

\subsection{Scaling form of covariance matrix}

The bridge between the continuum and lattice descriptions is that each lattice operator admits an expansion in terms of scaling operators \cite{Cardy1996},
\begin{equation}
    O_i(\mathbf{x})-\langle O_i\rangle \approx \sum_{a=1}^\infty W_{ai}\phi_a(\mathbf{x}), \qquad W_{ai}\in\mathbb C,
    \label{eq:lattice_expansion}
\end{equation}
where we assume we are working with $N_\ell$ lattice operators. We note that for a given lattice model at a particular fixed point the coefficients $W_{ai}$ although non-universal (model dependent) are well defined, and e.g. can in principle be determined from detailed data on the lattice statistical mechanics problem. 

An additional complication is that the correlation functions acquire a subleading correction in a lattice model at its critical temperature because while the system flows to the fixed point under RG, it deviates from it because of finite irrelevant operators \cite{Wegner1972,PelissettoVicari2002}. Corrections generated lead to an additive power law corrections at large distance, in which $k_{ab}\rightarrow K_{ab}(r)$
\begin{equation}
    K_{ab}(r) = k_{ab}+O\bigg(\frac{1}{r^{\omega}}\bigg), \qquad \omega>0,
    \label{eq:K_asymptotic}
\end{equation}
where $-\omega$ is the corresponding irrelevant RG eigenvalue. 

Putting these results together gives us a general form of the $N_\ell \times N_\ell$ operator covariance matrix at the critical point of a lattice model,
\begin{equation}
    C(r) = \sum_{a,b=1}^{\infty} \frac{K_{ab}(r)}{r^{\Delta_a+\Delta_b}} \ket{w_a}\bra{w_b}.
    \label{eq:full_covariance}
\end{equation}
where for notational convenience, we have collected the coefficients of each scaling operator in the lattice expansions into an $N_\ell$-component vector,
\begin{equation}
    \bra{w_a} = (W_{a1},\ldots,W_{aN_\ell}) \in\mathbb C^{N_\ell}, \qquad \ket{w_a}\equiv\bra{w_a}^{\dagger}.
    \label{eq:overlap_vector}
\end{equation}
Eq.~\ref{eq:full_covariance} is the general form for the covariance matrix we expect as a critical lattice model flows to a scale invariant fixed point, including corrections that arise from the flow to zero of the leading irrelevant variable. Away from very short distances of the order of a lattice spacing this form is expected to describe the lattice covariance matrix. 

\subsection{Linear Algebra Result}
\label{sec:linear_algebra_result}

Before inspecting the eigenvalues of the operator covariance matrix in Eq.~\eqref{eq:full_covariance}, we first state a general linear algebra result that we will make use of. Since we were unable to find this result published explicitly previously, we state and prove it in our manuscript. We first present a simpler version of the statement,a and then the full version we will need. In the main text we only give the statements, complete proofs are available in Appendix~\ref{app:determinant}.

 Let $G$ be a Hermitian $N\times N$ matrix which depends on a parameter $r$ and whose elements have the form,
\begin{equation}
    G_{ij} (r) = \frac{m_{ij}}{r^{\Omega_i+\Omega_j}},
    \label{eq:graded_matrix}
\end{equation}
where $m=(m_{ij})$ is a generic Hermitian $N\times N$ matrix satisfying the nonvanishing effective-block condition stated in Appendix~\ref{app:determinant}. The $N$ real exponents $\Omega_i>0$ are arranged in ascending order. For such a matrix, the asymptotic $r$-dependence of the eigenvalues of $G(r)$ as $r\rightarrow \infty$ is,
\begin{equation}
    \lambda_i(r)\sim c_i r^{-2\Omega_i}+ \dots, \qquad 1\leq i\leq N.
    \label{eq:main_result}
\end{equation}
This is a remarkably helpful result, since the exponents $\Omega_i$ that appear mixed up in the elements of $G(r)$, conveniently filter out in the leading behavior of the eigenvalues. We note that there are subleading corrections denoted by the ``$\dots$". 
The result holds even if there are finite degeneracies in the $\Omega_i$ - they will simply produce the corresponding number of eigenvalues with the same decay exponent. 

Now we are ready to state the more general form of the statement that we will make use of below. We can generalize the form of $G_{ij}(r)$ by allowing the matrix in the numerator of Eq.~\ref{eq:graded_matrix} to have $r$ dependence so long as it converges to a finite $m_{ij}$ when $r\rightarrow\infty$ and still have the result, Eq.~\ref{eq:main_result}, i.e. we can allow 
\begin{align}
        G_{ij} (r) = \frac{M_{ij}(r)}{r^{\Omega_i+\Omega_j}},\\
    {\rm with} \lim_{r\rightarrow\infty}M_{ij}(r)=m_{ij}.
    \label{eq:M_limit}
\end{align}
and Eq.\ref{eq:main_result} still holds, the only change will be in the subleading behavior, represented as ``$\dots$". Appendix~\ref{app:determinant} gives a direct proof of this result through solving the characteristic equation of the matrix $G(r)$.

\subsection{Operator Covariance}

In this subsection we show that using the result for the leading behavior of  eigenvalues allows us to prove that the eigenvalues of the general form of the critical covariance matrix,  Eq.~(\ref{eq:full_covariance}) also has the same scaling behavior. We show explicitly that, when the first $N_\ell$ overlap vectors are linearly independent and the limiting coefficient matrix satisfies the condition in Appendix~\ref{app:determinant}, the smallest $N_\ell$ scaling dimensions govern the leading behavior of the eigenvalues. Thus the set of arguments given below completes the proof of the result used by Sandvik.

There are essentially two steps we need to execute before we can apply the form Eq.~(\ref{eq:M_limit}) to Eq.~(\ref{eq:full_covariance}): First, we rewrite the infinite sum in Eq.~(\ref{eq:full_covariance}) using the first $N_\ell$ overlap vectors, retaining the higher-field contributions in the coefficient functions. Second, we orthogonalize this basis and show that the resulting matrix has the form in Eq.~(\ref{eq:M_limit}). We start with Eq.~(\ref{eq:full_covariance}) which is the form of the $N_\ell\times N_\ell$ covariance matrix for a lattice model at criticality. For clarity we reproduce it here,
\begin{equation*}
    C(r) = \sum_{a,b=1}^{\infty} \frac{K_{ab}(r)}{r^{\Delta_a+\Delta_b}} \ket{w_a}\bra{w_b}.
\end{equation*}
Notice there are infinite $|w_a\rangle$ vectors arranged by the values of the scaling dimensions, $\Delta_a$. On the other hand, the lattice operator space is finite $N_\ell$ dimensional. To cure this mismatch in the dimensions, assuming that $\ket{w_1},\ldots,\ket{w_{N_\ell}}$ are linearly independent, we use them to form a $N_\ell$-dimensional basis of the lattice-operator space. Hence, there is a unique rectangular matrix $S$, which has $N_\ell$ rows and infinitely many columns, such that
\begin{equation}
    \ket{w_b} = \sum_{a=1}^{N_{\ell}}S_{ab}\ket{w_a}.
    \label{eq:tail_expansion_app}
\end{equation}
Notice when $b \leq N_{\ell}$, $S_{ab}=\delta_{ab}$ since we retain $\{\ket{w_1},\ldots,\ket{w_{N_{\ell}}}\}$ as the basis at this step. Substituting Eq.~\eqref{eq:tail_expansion_app} into Eq.~\eqref{eq:full_covariance} gives
\begin{equation}
    C(r) = \sum_{a,b=1}^{N_{\ell}}  \frac{\overline{K}_{ab}(r)}{r^{\Delta_a+\Delta_b}} \ket{w_a}\bra{w_b},
    \label{eq:finite_covariance}
\end{equation}
where
\begin{equation}
    \overline{K}_{ab}(r)
    =\sum_{c,d\geq1}^{+\infty}S_{ac}S_{bd}^{*}K_{cd}(r) \frac{r^{\Delta_a+\Delta_b}}{r^{\Delta_c +\Delta_d}}.
    \label{eq:K_bar_app}
\end{equation}
We can then classify this sum into two parts: the first one involves the terms with the indices $1 \leq c,d\leq N_\ell$, which therefore yields nothing but $K_{ab}(r)$ since $S_{ac}=\delta_{ac}$ and $S_{bd}=\delta_{bd}$. The second part is composed of the terms with $c>N_\ell$ or $d>N_\ell$, and by the assumption that the spectrum of the scaling dimensions are sparse enough, the infinite sum will converge to the asymptotic limit. Hence, when $r \rightarrow +\infty$, every $\overline K_{ab}(r)$ converges to a finite constant, and $\overline{K}_{ab}(r)r^{-\Delta_a-\Delta_b}$ is still of order $r^{-\Delta_a-\Delta_b}$.

Up to now, the basis $\{\ket{w_1},\ldots,\ket{w_{N_\ell}}\}$ is still non-orthogonal. To obtain an orthonormal basis, we proceed to the second step: the application of the Gram--Schmidt process to $\ket{w_1},\ldots,\ket{w_{N_\ell}}$ in increasing order of the subscripts. We therefore derive the orthonormal basis $\{\ket{e_1},\ldots,\ket{e_{N_\ell}}\}$, and they are related to the original basis by
\begin{equation}
    \ket{w_a} = \sum_{b\leq a} \alpha_{ba}\ket{e_b}, \qquad \alpha_{ba} = \braket{e_b|w_a}.
    \label{eq:triangular_expansion_app}
\end{equation}
Substituting Eq.~\eqref{eq:triangular_expansion_app} into Eq.~\eqref{eq:finite_covariance} gives
\begin{equation}
    C(r) = \sum_{a,b=1}^{N_{\ell}} \frac{M_{ab}(r)}{r^{\Delta_a+\Delta_b}} \ket{e_a}\bra{e_b},
    \label{eq:graded_covariance_app}
\end{equation}
where
\begin{equation}
    M_{ab}(r)
    =\sum_{\substack{a\leq c\leq N_{\ell}\\b\leq d\leq N_{\ell}}}\alpha_{ac}\alpha_{bd}^{*}\overline{K}_{cd}(r) \frac{r^{\Delta_a+\Delta_b}}
    {r^{\Delta_c+\Delta_d}}.
    \label{eq:M_after_GS_app}
\end{equation}
Since $c\geq a$ and $d\geq b$, both exponent differences are non-negative. Therefore, terms with $\Delta_c>\Delta_a$ or $\Delta_d>\Delta_b$ vanish at large $r$, whereas mixing among equal dimensions remains finite. As results, every coefficient has a finite limit:
\begin{equation}
    \lim_{r\rightarrow+\infty} M_{ab}(r) = m_{ab}.
    \label{eq:M_after_GS_limit_app}
\end{equation}
Notice that Gram--Schmidt changes the coefficient functions but preserves the power $r^{-\Delta_a-\Delta_b}$ associated with every matrix element. Hence,  we have completed the reduction from a general operator covariance matrix in Eq.~\ref{eq:full_covariance} to the form in Eq.~\eqref{eq:M_limit}, with $N=N_\ell$ and $\Omega_a=\Delta_a$, and the asymptotic behavior of the eigenvalues in Eq.~\ref{eq:main_result} holds.

\section{Covariance eigenvectors and two-point amplitudes, $k_{ab}$}
\label{sec:covariance_eigenvectors_theory}

We have shown that the leading behavior of each eigenvalue is controlled by the scaling dimension $\Delta_a$ of a scaling operator $\phi_a$. Is it then correct to assume that the associated eigenvector $|u_a(r)\rangle$ (as the separation of the covariance matrix $r$ gets large) can provide us with a linear combination of lattice operators which is an approximation for the scaling operator $\phi_a$? We note firstly that since $\langle u_b (r)|u_a(r)\rangle = \delta_{ab}$ it is clearly incorrect to associate $|u_a(r)\rangle$ naively with scaling operators $\phi_a (r)$, since the scaling operators can have off-diagonal correlation functions, i.e. $k_{ab}\neq 0$ in general. We show nonetheless that the eigenvectors do contain valuable information and by a careful analysis which we propose here, the two-point amplitudes $k_{ab}$ in Eq.~\eqref{eq:continuum_correlator} can be extracted from the covariance matrix eigenvectors.

To start, we introduce a reference separation $R$ and recall the unitary matrix $U(R)$ that diagonalizes the covariance matrix $C(R)=U(R)D(R)U^{\dagger}(R)$. We denote its $a$-th column (i.e. the $a$-th eigenvector) by $\ket{u_a(R)}=\bigl(U_{1a}(R),\ldots,U_{N_\ell a}(R)\bigr)^{\mathsf T}$. As will prove useful in our proposed method, we use the eigenvector to construct the ``orthogonal'' lattice operator
\begin{equation}
    \widetilde O_a(r;R)=\sum_{i=1}^{N_\ell}U_{ia}(R)\bigl[O_i(r)-\langle O_i\rangle\bigr],
    \label{eq:eigenoperator}
\end{equation}
and study the cross correlation between them,
\begin{equation}
    \mathcal C_{ab}(r;R)=\left\langle\widetilde O_a^\dagger(0;R)\widetilde O_b(r;R)\right\rangle = [U^\dagger(R)C(r)U(R)]_{ab}.
    \label{eq:projected_covariance}
\end{equation}
For clarity, we call $R$ the construction distance and $r$ the measurement distance as motivated by Eq.~\eqref{eq:eigenoperator} and \eqref{eq:projected_covariance}.

Clearly while
\begin{equation}
 \mathcal C_{ab}(R;R)
    =\lambda_a(R)\delta_{ab},
    \label{eq:eigenoperator_orthogonality}
\end{equation}
 the off-diagonal elements of  $\mathcal C_{ab}(r;R)$ are non-zero for a general $r\neq R$. Indeed we will show that we can extract $k_{ab}$ from a careful analysis of the off-diagonal elements of $\mathcal C_{ab}(r;R)$.

We first study the form $\mathcal C_{ab}(r;R)$ constrained by scaling, by expanding Eq.~\eqref{eq:eigenoperator} in terms of the scaling operators. 
\begin{equation}
    \begin{aligned}
        \widetilde O_a(r;R)
        &=\sum_{c=1}^{+\infty}\braket{w_c|u_a(R)}\phi_c(r) \\
        &=\sum_{\Delta_c<\Delta_a}\frac{A_{ca}(R)}{R^{\Delta_a-\Delta_c}}\phi_c(r)
        +\sum_{\Delta_c\geq\Delta_a}q_{ca}(R)\phi_c(r).
    \end{aligned}
    \label{eq:eigenoperator_in_scaling_operator_expression}
\end{equation}
where in the second line we have divided the sum of scaling operators $\phi_c$, into two parts: the first term has the scaling operators with scaling dimensions $\Delta_c < \Delta_a$ and the second term has the operators $\Delta_c \geq \Delta_a$. Importantly, both coefficient functions $A_{ca}(R)$ and $q_{ca}(R)$ have finite magnitudes in large-$R$ limits, and generically $q_{aa}(R) \neq 0$. In Appendix~\ref{app:eigenoperator_asymptotics}, we explicitly derive Eq.~\eqref{eq:eigenoperator_in_scaling_operator_expression}. 

Equation~\eqref{eq:eigenoperator_in_scaling_operator_expression}  reveals an important property. At fixed $r$, the first sum vanishes in the limit $R\rightarrow+\infty$. In this limit  the leading behavior of $C_{ab}(r;R)$ will then be controlled by the correlation function of $\phi_a(r)$ and $\phi_b(r)$. We can pick up the normalized version of the $k_{ab}$ term by dividing by the diagonal correlations. This observation motivates us to consider the normalized cross ratio,
\begin{equation}
    \rho_{ab}(r;R)=\frac{\mathcal C_{ab}(r;R)}{\sqrt{|\mathcal C_{aa}(r;R)\mathcal C_{bb}(r;R)|}}.
    \label{eq:rho_definition}
\end{equation}
It is then apparent that, given $1 \leq a,b \leq N_{\ell}$, if $\phi_a$ and $\phi_b$ have \textit{non-degenerate} scaling dimensions,
\begin{equation}
    \lim_{r\rightarrow+\infty}
    \left[\lim_{R\rightarrow+\infty}|\rho_{ab}(r;R)|\right]
    =|k_{ab}|.
\label{eq:rho_scaling_limit}
\end{equation}
We here choose to focus first on the magnitude of $\rho_{ab}(r;R)$, since its individual phase is not unique. This non-uniqueness arises from re-phasing a covariance eigenvector according to $\ket{u_a(R)}\mapsto e^{i\psi_a}\ket{u_a(R)}$. Under this transformation, the normalized correlation changes as $\rho_{ab}(r;R)\mapsto e^{i(\psi_b-\psi_a)}\rho_{ab}(r;R)$. 

\begin{figure*}[t]
    \centering
    \includegraphics[width=\textwidth]{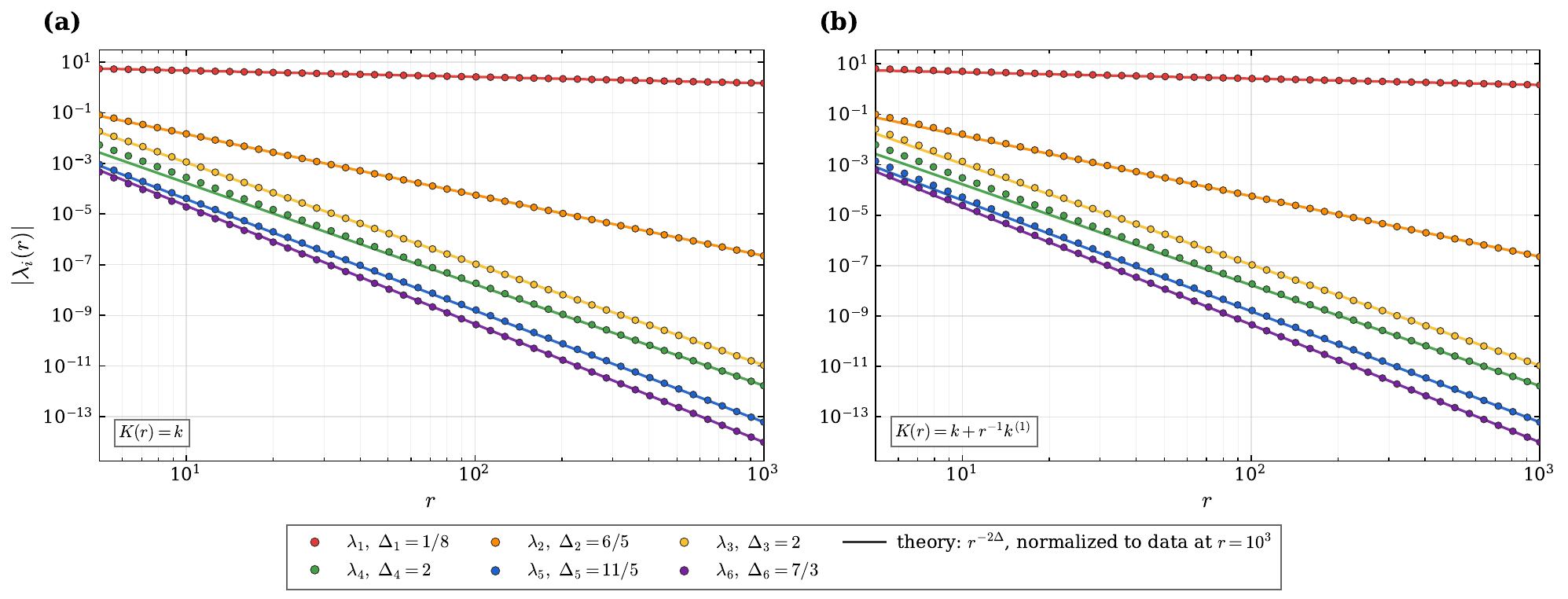}
    \caption{
        Synthetic demonstration of the covariance-eigenvalue scaling predicted by Eq.~\eqref{eq:main_result}. The dots show the absolute values of the six eigenvalues $\lambda_a(r)$. The solid comparison lines have the theoretical slopes $-2\Delta_a$ and pass through the corresponding data at $r=10^3$; they are not fits. They convincingly show that the eigenvalues of the covariance matrix decay with an exponent equal to the leading scaling dimensions. We have studied two different synthetic data set models as discussed in the text:  (a) Unperturbed coefficient matrix $K_{ab}(r)=k_{ab}$. (b) Coefficient matrix $K_{ab}(r)=k_{ab}+k^{(1)}_{ab}/r$. The matrix elements $W_{ab}$ and $k_{ab}$ are identical in the two panels, while $k^{(1)}_{ab}$ are independent matrix elements, which introduces a term to mimic the correction to the leading scaling behavior from an irrelevant operator. The irrelevant correction in (b) produces slightly stronger finite-distance curvature, but the branches approach the same theoretical powers eventually.
    }
    \label{fig:synthetic_eigenvalues}
\end{figure*}


We highlight that a most interesting feature that $|k_{ab}|$ can diagnose: providing evidence for conformal invariance at criticality. At most critical points, scale invariance is enhanced to conformal invariance \cite{Cardy1984}, which imposes its own constraints \cite{BelavinPolyakovZamolodchikov1984,FriedanQiuShenker1984,DiFrancesco1997,Rychkov2016,SimmonsDuffin2016}. One example is the orthogonality between conformal towers predicted by conformal field theory (CFT): scaling operators belonging to distinct conformal towers have vanishing two-point correlations \cite{DiFrancesco1997,Rychkov2016,SimmonsDuffin2016}. Hence, if the normalized cross ratio converges to zero in the ordered limits of Eq.~\ref{eq:rho_scaling_limit}, it suggests that the corresponding scaling operators are orthogonal, providing an indication of conformal invariance at the critical point.

In analyzing the limit, Eq.~\ref{eq:rho_scaling_limit}, from numerical data with various $r$ and $R$, we note that there is a simple scaling property,
\begin{equation}
\mathcal \rho_{ab}(r;R) = \Phi_{ab} \left (\frac{r}{R}\right)+\ldots,
\label{eq:scaling_form_of_projected_covariance}
\end{equation}
where $\ldots$ vanishes in the large-$R$ limit. We show this in the Appendix~\ref{app:eigenoperator_asymptotics} and check it numerically in the next two sections. In this scaling form, $k_{ab}=\Phi_{ab}(0^+)$. From Eq.~\eqref{eq:scaling_form_of_projected_covariance}, it is then convenient to plot $|\rho_{ab}(r;R)|$ against $r/R$ for different $R$, and the scaling collapse of these curves, together with the trend of $r/R \rightarrow 0^+$, can therefore be utilized to extract $|k_{ab}|$. An important case is $k_{ab}=0$, e.g. where the orthogonality between the scaling operators comes from conformal invariance. In this case, if the two resolved eigenvector branches $a$ and $b$ correspond to scaling operators belonging to distinct conformal towers, we have $\Phi_{ab}(r/R)=0$ for any $0<r/R\leq 1$, as shown in Appendix~\ref{app:cross_tower_collapse}. Consequently, the direct scaling collapse of $\rho_{ab}(r;R)$ fails. Nevertheless, if the leading non-vanishing correction is of order $R^{-g}$ with $g>0$, the rescaled correlation $R^g|\rho_{ab}(r;R)|$ can instead collapse onto a nontrivial function of $r/R$. We verify this numerically as well.

\section{Numerical Verification with Synthetic Data}
\label{sec:numerics}

We now provide tests of our central results the asymptotic predictions in Eq.~\eqref{eq:main_result} and in Eq.~\eqref{eq:rho_scaling_limit}.  In this section, rather than compute the covariance matrices from a statistical mechanics problem (we will do that in Sec.~\ref{sec:ising_application}), here we generate the covariance matrices from the scaling form in Eq.~\eqref{eq:full_covariance} under fairly general conditions: we take the normalized two-point amplitude matrix $k_{ab}$ in Eq.~\eqref{eq:continuum_correlator} to be non-diagonal and the complex overlap vectors $\ket{w_a}$ to be non-orthogonal. We also consider a spectrum with degenerate scaling dimensions and deliberately choose the number $N_\ell$ of measured lattice operators to be smaller than the number of continuum operators, demonstrating that truncating the operator basis at $N_\ell$ does not alter the asymptotic scaling of the resolved eigenvalues. In Sec.~\ref{ssec:irr}, we further introduce an irrelevant perturbation of the form given in Eq.~\eqref{eq:K_asymptotic} and show that the same asymptotic powers are recovered. Finally, in Sec.~\ref{ssec:synthetic_eigenvectors}, we show how $|k_{ab}|$ can be extracted from the normalized cross ratio.

\subsection{Synthetic data without irrelevant perturbations}

To generate synthetic data, we use six lattice operators and nine continuum scaling operators with dimensions
\begin{equation}
    \left( \frac{1}{8}, \frac{6}{5}, 2, 2, \frac{11}{5}, \frac{7}{3}, \frac{7}{3}, \frac{7}{3}, 3 \right).
    \label{eq:synthetic_dimensions}
\end{equation}
As for the expansion coefficients $W_{ai}$ introduced in Eq.~\eqref{eq:lattice_expansion}, we draw them from the complex Gaussian ensemble,
\begin{equation}
    W_{ai}=z_{ai}=\frac{x_{ai}+\mathrm{i}y_{ai}}{\sqrt{2}},
    \label{eq:synthetic_W}
\end{equation}
where $x_{ai}$ and $y_{ai}$ are independent standard normal variables. Thus, $\mathbb{E}|z_{ai}|^2=1$. We impose neither normalization nor orthogonality on the overlap vectors $\ket{w_a}$. Hence, the first six rows are almost surely linearly independent. All nine rows contribute to the covariance matrix, and the last three therefore provide a finite proxy for the unresolved tail of continuum operators.

Independently, we generate a Hermitian normalized two-point amplitude matrix according to
\begin{equation}
    k_{aa}=1, \qquad
    k_{ab}=\frac{1}{4}\xi_{ab}, \qquad
    k_{ba}=k_{ab}^{*},
    \label{eq:synthetic_K}
\end{equation}
for every $a<b$, where the $\xi_{ab}$ are independent standard complex Gaussian variables with $\mathbb{E}|\xi_{ab}|^2=1$. The off-diagonal entries are therefore random, with root-mean-square magnitude $1/4$.

Upon initialization of $k$ and the overlap vectors $\ket{w_a}$, we evaluate the operator covariance matrix at logarithmically spaced separations between $r=5$ and $r=10^3$. We label the eigenvalues $\lambda_1(r),\ldots,\lambda_6(r)$ according to their asymptotic powers and follow the branches continuously between adjacent separations. For comparison, we draw straight lines with the theoretical slopes $-2\Delta_a$ through the corresponding eigenvalue data at the largest separation, $r=10^3$. These lines are not fits. Figure~\ref{fig:synthetic_eigenvalues}(a) demonstrates how the six eigenvalue branches approach the powers associated with $\Delta_1=1/8$, $\Delta_2=6/5$, $\Delta_3=\Delta_4=2$, $\Delta_5=11/5$, and $\Delta_6=7/3$. In particular, $\lambda_3$ and $\lambda_4$ approach distinct curves but with the same exponent because $\Delta_3=\Delta_4=2$. Since $k$ is independent of $r$, the deviations at shorter distances arise from mixing among the scaling contributions. Overall, the asymptotic behavior of the eigenvalues agrees well with the prediction from Eq.~\eqref{eq:main_result}.

\subsection{Synthetic data with an irrelevant perturbation}
\label{ssec:irr}
We next retain the same realizations of $W_{ab}$ and $k_{ab}$ but replace the constant coefficient matrix by
\begin{equation}
    K_{ab}(r) = k_{ab}+\frac{k^{(1)}_{ab}}{r}.
    \label{eq:synthetic_irrelevant_K}
\end{equation}
The newly-added matrix $k^{(1)}_{ab}$, which mimics the irrelevant perturbation in Eq.~\eqref{eq:K_asymptotic}, is drawn independently from a Hermitian ensemble,
\begin{equation}
    k^{(1)}_{aa}=1, \qquad
    k^{(1)}_{ab}=\frac{1}{2}\zeta_{ab}, \qquad
    k^{(1)}_{ba}=\bigl(k^{(1)}_{ab}\bigr)^{*},
    \label{eq:synthetic_K1}
\end{equation}
for $a<b$, where the $\zeta_{ab}$ are independent complex Gaussian variables. The off-diagonal entries of $k^{(1)}$ therefore have a root-mean-square magnitude of $1/2$. Equation~\eqref{eq:synthetic_irrelevant_K} corresponds to an irrelevant amplitude $1$ and exponent $\omega=1$ in Eq.~\eqref{eq:K_asymptotic}, or RG eigenvalue $-1$. In particular, the off-diagonal correction remains generic and is independent from $k$ and $W$.

As result, the irrelevant perturbation changes the finite-distance amplitudes but not the asymptotic powers. One contribution to a generic eigenvalue still has the form
\begin{equation}
    \lambda_a(r) = \kappa_a r^{-2\Delta_a} \left[ 1+\ldots \right],
    \label{eq:synthetic_irrelevant_eigenvalue}
\end{equation}
where ``...'' vanishes in the large-$r$ limit. Figure~\ref{fig:synthetic_eigenvalues}(b) shows the resulting spectrum over the same range $5\leq r\leq10^3$. For relatively small $r<10$, the irrelevant perturbation produces slightly greater deviation from the theoretical reference lines in comparison with the unperturbed case, while all the eigenvalue branches nevertheless quickly bend toward the same powers as $r$ increases. The comparison between the two panels demonstrates how an irrelevant perturbation can delay the onset of the asymptotic regime without changing its behavior.

\begin{figure}[t]
    \centering
    \includegraphics[width=\columnwidth]{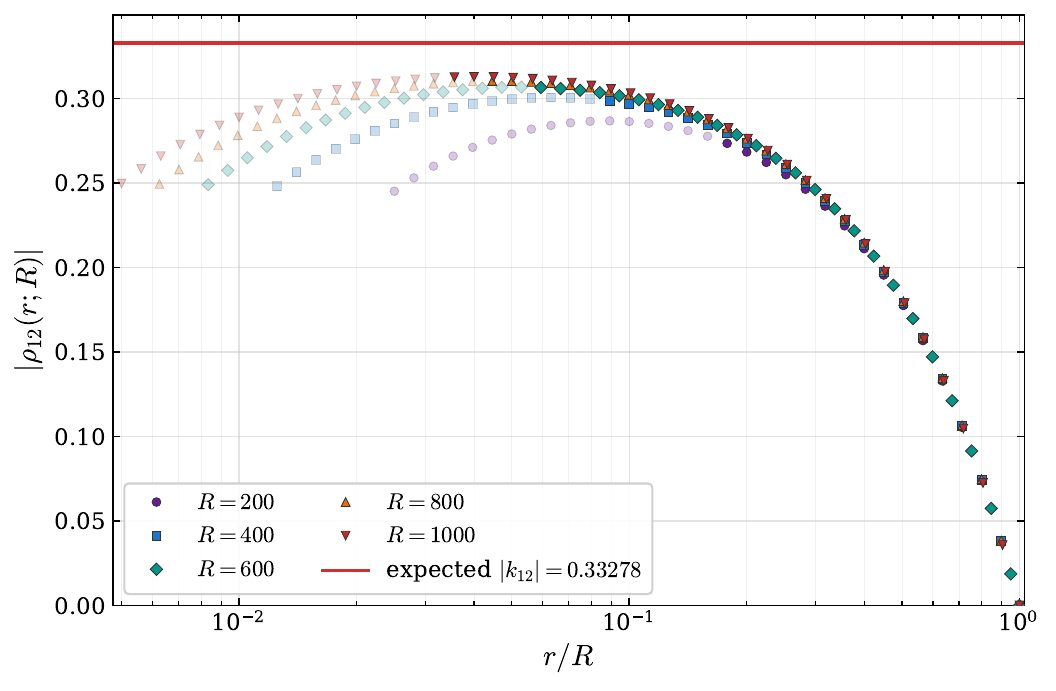}
    \caption{
        Magnitude $|\rho_{12}(r;R)|$ of the normalized cross ratio for the first two eigenvectors of the perturbed complex synthetic covariance matrix, plotted against the relative separation $r/R$. The eigenvectors are frozen at $R=200$, $400$, $600$, $800$, and $1000$, and each curve is evaluated over $5\leq r\leq R$. The horizontal red line marks the expected value $|k_{12}|=0.33278\ldots$ from the random input. Every ratio vanishes at $r/R=1$ because the corresponding eigenvectors diagonalize $C(R)$. Points with $r<32$ are shown in lighter shades to distinguish the short-distance region; they are not omitted from the plot. With increasing $R$, the data approach one another over intermediate relative separations and move toward the expected value.
    }
    \label{fig:synthetic_eigenvectors}
\end{figure}

\subsection{Synthetic test of covariance eigenvectors}
\label{ssec:synthetic_eigenvectors}

We now test Eq.~\eqref{eq:rho_scaling_limit} using the first two eigenvectors of the perturbed synthetic covariance matrix. The two target dimensions, $\Delta_1=1/8$ and $\Delta_2=6/5$, are nondegenerate. For the random realization chosen in Sec.~\ref{ssec:irr}, the expected magnitude of corresponding the two-point amplitude is
\begin{equation}
    |k_{12}|=0.332782\ldots.
    \label{eq:synthetic_rho_scaling}
\end{equation}
Hence, from Eq.~\eqref{eq:rho_scaling_limit}, we anticipate
\begin{equation}
    \lim_{r \rightarrow +\infty}\left[\lim_{R \rightarrow +\infty}|\rho_{12}(r;R)|\right] = |k_{12}|
    \label{eq:synthetic_two_point_amplitude}
\end{equation}
To demonstrate this, in Fig.~\ref{fig:synthetic_eigenvectors}, we measure $|\rho_{12}(r;R)|$ while varying $r/R$ for $R=200$, $400$, $600$, $800$, and $1000$, with each curve evaluated over $5\leq r\leq R$. Plotting against $r/R$ places the exact zero of every curve at the common point $r/R=1$. Moreover, it demonstrates the scaling collapse between curves with different $R$, hence allowing the results at different construction distances to be compared directly. As $R$ increases, the collapsed range of $|\rho_{12}(r;R)|$ extends, and the collapsed curve approaches the expected value in Eq.~\eqref{eq:synthetic_rho_scaling} when $r/R$ decreases. At $R=1000$, the largest sampled value is $0.3128\ldots$ at $r/R=0.04$, about $6\%$ below $|k_{12}|$. Thus, the data $|\rho_{12}(r;R)|$, while being $0$ at the measurement distance $r=R$, display a trend toward the scaling-field correlation under the double limit $\lim_{r/R \rightarrow 0^+}\lim_{R \rightarrow +\infty}$. This illustrates how the scaling collapse of $|\rho_{ab}(r;R)|$ against $r/R$ can extract nonzero $|k_{ab}|$ numerically.

\section{Application to the Two-Dimensional Ising Model}
\label{sec:ising_application}

Now that we have tested the key statements in Eqs.~\eqref{eq:main_result} and \eqref{eq:rho_scaling_limit} using synthetic data, we apply the operator covariance method to the critical two-dimensional Ising model \cite{Onsager1944,McCoyWu1973,Wu1982}, whose Hamiltonian is
\begin{equation}
    \beta H=-K_c\sum_{\langle\mathbf x,\mathbf y\rangle}
    S_{\mathbf x}S_{\mathbf y},\qquad S_{\mathbf x}\in\{\pm1\},
    \label{eq:ising_hamiltonian}
\end{equation}
where $\langle\mathbf x,\mathbf y\rangle$ denotes pairs of nearest-neighbor sites and $K_c$ is the critical dimensionless coupling. We will study both square and triangular lattices since they allow us to explore different ranges of conformal spin. The triangular results are presented in this section and the square results are in Appendix~\ref{app:square}.

Although Ref.~\cite{Sandvik2026} applied the operator covariance method to the square lattice model, its Monte Carlo data resolved only a limited number of scaling dimensions because of statistical uncertainties. To comprehensively examine the capabilities of the operator covariance method, here we use numerically exact lattice-operator correlations derived on the infinite lattice. These correlations can be evaluated through the Pfaffian representation of the Ising model \cite{Kaufman1949,KacWard1952,Kasteleyn1963,Fisher1966,ChelkakCimasoniKassel2017}. In this representation, multispin correlation functions reduce to finite-dimensional matrix calculations \cite{MontrollPottsWard1963}. Constructing these matrices from the infinite-lattice Green's function appropriate to each geometry lets us evaluate the correlations directly in the thermodynamic limit with controlled numerical precision \cite{ThomasMiddleton2013,Cserti2000,Zhugayevych2025}. We use 256-bit arithmetic, approximately $256\log_{10}2\approx77$ decimal digits, to control round-off errors in the smallest covariance eigenvalues. This allows us to examine how much of the continuum operator spectrum can be extracted from a set of local spin products. We used OpenAI Codex (GPT-5.6 Sol) to assist with code development and debugging. The AI-assisted implementations were reviewed by the authors and validated against examples obtained from an independent computational pipeline based on Toeplitz determinants \cite{Stephenson1964,ChengWu1967}.

\begin{table*}[t]
\caption{Point sets used to construct the even-sector operators. The last column lists the lattice spins $\tilde{s}$ for which the operator $P_{\tilde{s}}E(\mathbf x;X_{\mu})$ is nonzero.}
\label{tab:triangular_energy_lattice_operators}
\begin{ruledtabular}
\begin{tabular}{cll}
 & points in $X_{\mu}$ & allowed lattice spins $\tilde{s}$ \\
\hline
$X_1$ & $\{\mathbf0,\mathbf e_1\}$ & $0,1,2,3,4,5$ \\
$X_2$ & $\{\mathbf0,\mathbf e_1+\mathbf e_2\}$ & $0,1,2,3,4,5$ \\
$X_3$ & $\{-\mathbf e_1,\mathbf e_1\}$ & $0,2,4$ \\
$X_4$ & $\{-\mathbf e_1-\mathbf e_2,\mathbf e_1+\mathbf e_2\}$ & $0,2,4$ \\
$X_5$ & $\{\mathbf0,\mathbf e_1,\mathbf e_2,\mathbf e_1+\mathbf e_2\}$ & $0,1,2,3,4,5$ \\
$X_6$ & $\{\mathbf e_1,\mathbf e_2,-\mathbf e_1+\mathbf e_2,-\mathbf e_1,-\mathbf e_2,\mathbf e_1-\mathbf e_2\}$ & $0$ \\
$X_7$ & $\{\mathbf e_1+\mathbf e_2,-\mathbf e_1+2\mathbf e_2,-2\mathbf e_1+\mathbf e_2,-\mathbf e_1-\mathbf e_2,\mathbf e_1-2\mathbf e_2,2\mathbf e_1-\mathbf e_2\}$ & $0$ \\
\end{tabular}
\end{ruledtabular}
\end{table*}

To interpret the resulting spectrum, it is useful to review the two-dimensional Ising conformal field theory (CFT), which describes the critical Ising universality class \cite{BelavinPolyakovZamolodchikov1984,FriedanQiuShenker1984,Ginsparg1988,DiFrancesco1997,BlumenhagenPlauschinn2009,ChelkakHonglerIzyurov2015}. The theory contains three Virasoro primary fields: the identity $\mathbb I$, the spin field $\sigma$, and the energy field $\varepsilon$. All three have zero conformal spin, with scaling dimensions $0$, $1/8$, and $1$, respectively. More generally, descendants of a primary $\phi\in\{\mathbb I,\sigma,\varepsilon\}$ are generated by the Virasoro operators $L_{-n}$ and $\bar L_{-\bar n}$ and take the form
\begin{equation}
    L_{-n_1}\cdots L_{-n_p}
    \bar L_{-\bar n_1}\cdots\bar L_{-\bar n_q}\phi.
    \label{eq:ising_descendant_form}
\end{equation}
For holomorphic and antiholomorphic levels $\ell=\sum_i n_i$ and $\bar\ell=\sum_i\bar n_i$, their scaling dimension and conformal spin are
\begin{equation}
    \Delta=\Delta_\phi+\ell+\bar\ell,\qquad s=\ell-\bar\ell.
    \label{eq:ising_descendant_quantum_numbers}
\end{equation}
Thus, a field (equivalently a state), which is either a primary or a descendant, can generically be classified by the pair of numbers $(\Delta,s)$.

Some of these descendants have a familiar interpretation as spatial derivatives: $L_{-1}\phi=\partial\phi$ and $\bar L_{-1}\phi=\bar\partial\phi$ \cite{Ginsparg1988,DiFrancesco1997,BlumenhagenPlauschinn2009}. Repeated derivatives therefore generate part of each conformal tower. However, a descendant in the $2$D CFT need not be obtained merely by differentiating the primary. For example, $T=L_{-2}\mathbb I$ is the holomorphic component of the stress-energy tensor, even though every derivative of the identity itself vanishes. Resolving such fields tests whether the covariance method can reveal operator content beyond derivatives of the leading spin and energy primaries.

Not all expressions in Eq.~\eqref{eq:ising_descendant_form} represent independent states, because the Ising Verma modules contain null states \cite{BelavinPolyakovZamolodchikov1984,FriedanQiuShenker1984,RochaCaridiWallach1984,Ginsparg1988,DiFrancesco1997}. These relations must be removed when counting multiplicities. They can also identify a Virasoro descendant with a derivative; for example, $L_{-2}\sigma$ is proportional to $L_{-1}^2\sigma$. Thus, the occurrence of a generator $L_{-n}$ with $n>1$ alone does not establish an additional independent field. In the tables below, we list convenient representatives after accounting for the null relations. Several states in one row describe the allowed CFT subspace, rather than a unique assignment of each covariance eigenvector to a particular representative.

For the triangular lattice, we tune the coupling to its critical value $K_c=\frac14\ln3$ \cite{Wu1982}. The different rotational symmetries of the two lattices organize the same continuum spectrum into different lattice-spin sectors.

\subsection{Triangular lattice}
\label{sec:ising_triangular_lattice}

We start with the construction of the lattice operators. Given a finite set $X$ of lattice vectors, we define the local lattice operator at site $\mathbf x$ by
\begin{equation}
    E(\mathbf x;X)=\prod_{\boldsymbol\delta\in X}S_{\mathbf x+\boldsymbol\delta}
    -\left\langle\prod_{\boldsymbol\delta\in X}S_{\mathbf x+\boldsymbol\delta}\right\rangle.
    \label{eq:ising_lattice_operator}
\end{equation}
The set $X$ specifies the arrangement of spins in the product, while changing $\mathbf x$ translates the whole operator. Subtracting the expectation value removes the identity. Notice that it does not remove nontrivial descendants in the identity tower though, such as the stress-energy tensor, which can still contribute to correlations.

Since the triangular lattice has sixfold rotational symmetry, we introduce the $(2\pi/6)$-rotation $Q_6$ on the plane and define the $q$-fold rotation of the set $X$ by
\begin{equation}
    Q_6^q[X]=\{Q_6^q\boldsymbol\delta:\boldsymbol\delta\in X\}.
    \label{eq:triangular_rotated_point_set}
\end{equation}
This means that we rotate every vector in $X$ by the same angle $2\pi q/6$, keeping the reference site $\mathbf x$ fixed. Replacing $X$ by $Q_6^q[X]$ in Eq.~\eqref{eq:ising_lattice_operator} then gives the same local spin product in a different orientation. We combine these orientations to construct an operator with the lattice spin $\tilde{s}$, distinguished from the continuum conformal spin $s$:
\begin{equation}
    P_{\tilde{s}}E(\mathbf x;X)=\frac{\sqrt{\ell_X}}{6}\sum_{q=0}^{5}
    e^{-2\pi i \tilde{s} q/6}E(\mathbf x;Q_6^q[X]),
    \label{eq:triangular_spin_projector}
\end{equation}
where $\tilde{s}\in\mathbb Z/6\mathbb Z$ and $\ell_X$ is the number of distinct point sets obtained by rotating $X$. This normalization gives each nonzero operator a coefficient vector of unit norm over the distinct orientations. The phase factors select how the operator transforms under a rotation by $2\pi/6$. Continuum fields with spins differing by six acquire the same phase under this rotation, so they belong to the same lattice-spin sector. That is to say, fields with conformal spins $s=\tilde{s}+6n$, where $n\in\mathbb Z$, all occur in the sector $\tilde{s}$.

To construct the covariance matrix, we first correlate two lattice operators with the same lattice spin $\tilde{s}$ along a fixed direction $\hat{\mathbf r}=\mathbf{r}/r$ and define
\begin{equation}
    \widetilde C_{\mu\nu}^{(\tilde{s})}(\mathbf r)
    =\left\langle P_{\tilde{s}}E(\mathbf0;X_{\mu})P_{\tilde{s}}E(\mathbf r;X_{\nu})\right\rangle. \label{eq:bilinear_correlation_of_lattice_operators}
\end{equation}
Notice that this bilinear correlation $\widetilde C_{\mu\nu}^{(\tilde{s})}(\mathbf r)$ carries lattice spin $2\tilde{s}$, so $\widetilde C_{\mu\nu}^{(\tilde{s})}$ may not be Hermitian when $\tilde{s} \neq 0$. Moreover, for spinful continuum fields, Eq.~\eqref{eq:bilinear_correlation_of_lattice_operators} has direction-dependent two-point amplitudes $f_{ab}(\hat{\mathbf r})$ \cite{Cardy1996}. However, along the fixed direction $\hat{\mathbf{r}}$, these amplitudes, including their phases, are constants \cite{Cardy1996}. Therefore, we have the constant argument 
\begin{equation}
     \chi_{\mu,\tilde{s}}=\frac12\arg\!\left[\widetilde C_{\mu\mu}^{(\tilde{s})}(\mathbf r_0)\right].
    \label{eq:phase_of_bilnear_correlation}
\end{equation}
Here $\mathbf r_0=r_0\hat{\mathbf{r}}$ is an arbitrary fixed reference separation as long as the lattice operators do not overlap. As results, we construct the operator covariance matrix
\begin{equation}
    C_{\mu\nu}^{(\tilde{s})}(\mathbf r)
    =e^{-i\chi_{\mu,\tilde{s}}-i\chi_{\nu,\tilde{s}}}\left\langle P_{\tilde{s}}E(\mathbf0;X_{\mu})P_{\tilde{s}}E(\mathbf r;X_{\nu})\right\rangle,
    \label{eq:triangular_complex_covariance}
\end{equation}
which is then guaranteed to be Hermitian. Also, these correlations are connected because the operators in Eq.~\eqref{eq:ising_lattice_operator} are centered. Hence, the matrix analysis of Sec.~\ref{sec:setup}, which yields the asymptotic behavior of the eigenvalues with the decay power-law associated with the scaling dimensions, applies.

We further classify the lattice operators according to their transformation under the global Ising spin flip. Products containing an even number of spins are invariant under this transformation and can couple to the $\varepsilon$ and $\mathbb I$ towers, forming what we call the even sector. In contrast, products containing an odd number of spins change sign under the spin flip and can couple to the $\sigma$ tower, forming the odd sector.

For the numerical analysis, we use integer separations $5\leq r\leq192$, in units of the nearest-neighbor spacing. The lower bound ensures that the two local lattice operators do not overlap. We analyze the absolute eigenvalues $|\lambda_a(r)|$. The high-precision values throughout $96\leq r\leq192$ are used to extract $\Delta_{\rm fit}$. The errors quoted in the tables are formal standard errors from the log--log linear regressions. For comparison, the figures also include straight reference lines with theoretical slopes $-2\Delta_a$, normalized to the last displayed point of each branch. The fitted dimensions are reported separately in the tables. All reported fitted dimensions agree with their assigned CFT values to within one percent, demonstrating consistency between the CFT spectrum and the infinite-lattice Ising correlations.

\begin{figure*}[t]
    \centering
    \includegraphics[width=\textwidth]{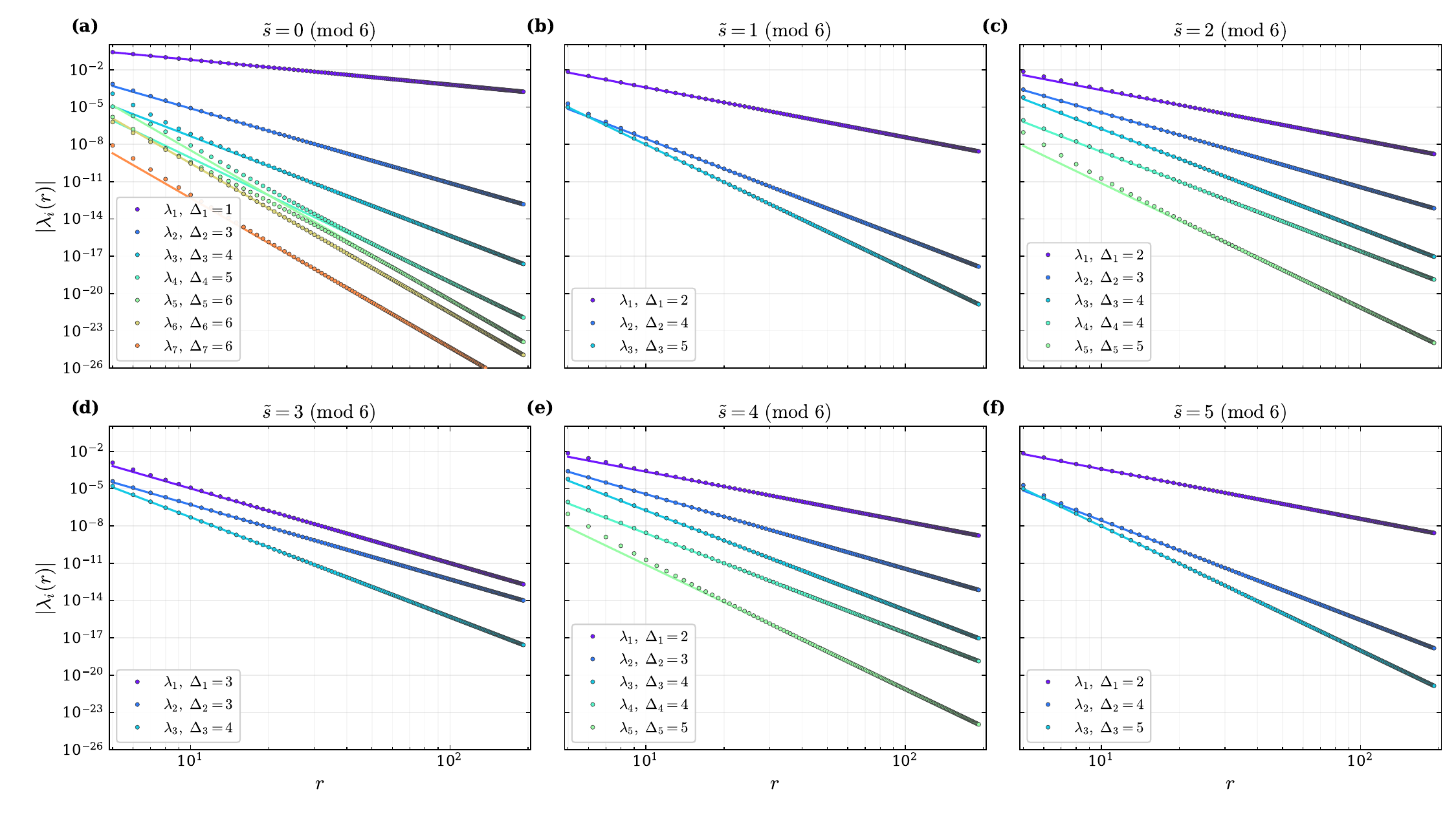}
    \caption{Even-sector covariance spectra on the infinite triangular lattice Ising model for $5\leq r\leq192$, obtained using the Pfaffian method as described in the text. Panels (a)--(f) show $\tilde{s}=0,1,\ldots,5\pmod{6}$. The dots are the numerically exact plane data with the display threshold at $10^{-26}$. The legends identify the eigenvalue branches and their theoretical dimensions. The solid line associated with each branch has the theoretical slope $-2\Delta$ and passes through its last displayed points; these lines are not fits. The fitted scaling dimensions from the data over the range $96 \leq r \leq 192$ are listed in Table.~\ref{tab:triangular_energy_c6}.}
    \label{fig:triangular_energy_c6}
\end{figure*}

\begin{table*}[t]
\caption{Even-sector CFT states associated with the resolved eigenvalue branches on the triangular lattice. The lattice spin $\tilde{s}$ is understood modulo six. Fitted dimensions are obtained from the data over $96\leq r\leq192$ in Fig.~\ref{fig:triangular_energy_c6}. Fitting errors (as parenthesized) are formal regression standard errors.}
\label{tab:triangular_energy_c6}
\begin{ruledtabular}
\begin{tabular}{ccll}
$\tilde{s}\ (\mathrm{mod}\ 6)$ & $\Delta$ & CFT states & $\Delta_{\rm fit}$ \\
\hline
$0$ & 1 & $\varepsilon$ & $1.0000424(8)$ \\
$0$ & 3 & $L_{-1}\bar L_{-1}\varepsilon$ & $3.000401(7)$ \\
$0$ & 4 & $L_{-2}\bar L_{-2}\mathbb I$ & $4.00170(3)$ \\
$0$ & 5 & $L_{-2}\bar L_{-2}\varepsilon$ & $5.0335(6)$ \\
$0$ & 6 & $L_{-6}\mathbb I,\ L_{-4}L_{-2}\mathbb I,\ L_{-3}^{2}\mathbb I,\ L_{-3}\bar L_{-3}\mathbb I,\ \bar L_{-6}\mathbb I,\ \bar L_{-4}\bar L_{-2}\mathbb I,\ \bar L_{-3}^{2}\mathbb I$ & $5.9676(6),\ 6.000099(2),\ 6.0056(1)$ \\
$1$ & 2 & $L_{-1}\varepsilon$ & $2.000183(3)$ \\
$1$ & 4 & $L_{-2}\bar L_{-1}\varepsilon$ & $4.00067(1)$ \\
$1$ & 5 & $L_{-3}\bar L_{-2}\mathbb I,\ \bar L_{-5}\mathbb I,\ \bar L_{-3}\bar L_{-2}\mathbb I$ & $5.000241(5)$ \\
$2$ & 2 & $L_{-2}\mathbb I$ & $2.00099(2)$ \\
$2$ & 3 & $L_{-2}\varepsilon$ & $2.999911(2)$ \\
$2$ & 4 & $\bar L_{-4}\mathbb I,\ \bar L_{-2}^{2}\mathbb I$ & $4.000416(8),\ 4.00105(2)$ \\
$2$ & 5 & $L_{-3}\bar L_{-1}\varepsilon,\ \bar L_{-4}\varepsilon,\ \bar L_{-3}\bar L_{-1}\varepsilon$ & $5.0065(1)$ \\
$3$ & 3 & $L_{-3}\mathbb I,\ \bar L_{-3}\mathbb I$ & $3.00093(2),\ 3.000174(3)$ \\
$3$ & 4 & $L_{-3}\varepsilon,\ \bar L_{-3}\varepsilon$ & $3.999825(3)$ \\
$4$ & 2 & $\bar L_{-2}\mathbb I$ & $2.00099(2)$ \\
$4$ & 3 & $\bar L_{-2}\varepsilon$ & $2.999911(2)$ \\
$4$ & 4 & $L_{-4}\mathbb I,\ L_{-2}^{2}\mathbb I$ & $4.000416(8),\ 4.00105(2)$ \\
$4$ & 5 & $L_{-4}\varepsilon,\ L_{-3}L_{-1}\varepsilon,\ L_{-1}\bar L_{-3}\varepsilon$ & $5.0065(1)$ \\
$5$ & 2 & $\bar L_{-1}\varepsilon$ & $2.000183(3)$ \\
$5$ & 4 & $L_{-1}\bar L_{-2}\varepsilon$ & $4.00067(1)$ \\
$5$ & 5 & $L_{-5}\mathbb I,\ L_{-3}L_{-2}\mathbb I,\ L_{-2}\bar L_{-3}\mathbb I$ & $5.000241(5)$ \\
\end{tabular}
\end{ruledtabular}
\end{table*}

\subsubsection{Even sector}
\label{sec:ising_energy_sector}

In the even sector, we use seven point sets $X_1,\ldots,X_7$ to construct the corresponding operators $P_{\tilde{s}}E(\mathbf x;X_{\mu})$. These sets are listed in Table~\ref{tab:triangular_energy_lattice_operators} using axial coordinates, where $\mathbf e_1=(1,0)$ and $\mathbf e_2=(1/2,\sqrt{3}/2)$ denote the two primitive lattice directions. We emphasize that there is no particular reason for choosing precisely these seven point sets. The point sets used in the other numerical examples in this manuscript are also chosen arbitrarily. This rather arbitrary choice actually demonstrates that the vectors $\ket{w_a}$, which contain the expansion coefficients of the lattice operators in terms of the scaling operators in Eqs.~\ref{eq:full_covariance} and \ref{eq:overlap_vector}, are generally linearly independent.

The six lattice-spin sectors have dimensions $(7,3,5,3,5,3)$ for $\tilde{s}=0,1,\ldots,5$, respectively, and account for 26 lattice operators. Figure~\ref{fig:triangular_energy_c6} shows the resulting eigenvalue spectra. We first examine the $\tilde{s}=0$ sector, whose seven branches approach dimensions $1$, $3$, $4$, $5$, and three copies of $6$. The first branch gives the scaling dimension of the energy primary $\varepsilon$. The second is consistent with $L_{-1}\bar L_{-1}\varepsilon$, which is proportional to the Laplacian of the energy field. This is the expected first step in separating a primary from its more rapidly decaying derivative contributions. Importantly, the branch at $\Delta=4$ shows that the method also resolves a field beyond these derivatives. Its CFT assignment is $L_{-2}\bar L_{-2}\mathbb I=T\bar T$, where $T$ and $\bar T$ are the holomorphic and antiholomorphic components of the stress-energy tensor \cite{DiFrancesco1997}. This field belongs to the identity tower and cannot be obtained by taking derivatives of the identity or of the energy primary. Thus, although the measured lattice operators are local products of Ising spins, their covariance matrix distinguishes contributions from different conformal towers in the same symmetry sector.

It is also intriguing to look into sectors with nonzero lattice spins as they contain non-trivial CFT states. For example, the stress-energy tensor itself appears at $\Delta=2$ in the $\tilde{s}=2$ and $\tilde{s}=4$ sectors, consistent with $T=L_{-2}\mathbb I$ and $\bar T=\bar L_{-2}\mathbb I$ \cite{DiFrancesco1997}. The remaining sectors contain further descendants of $\varepsilon$ and $\mathbb I$, with assignments listed in Table~\ref{tab:triangular_energy_c6}. Reflection symmetry is reflected in the coincident spectra of $\tilde{s}=1,5$ and $\tilde{s}=2,4$. Notice that at higher scaling dimensions, the number of independent lattice operators limits the resolved multiplicity. For example, in the scalar sector $\tilde{s}=0$, four of the seven eigenvalue branches already correspond to dimensions below six, leaving only three branches to represent the larger CFT subspace at $\Delta=6$. Its incomplete multiplicity is therefore expected for this finite operator basis.

Table~\ref{tab:triangular_energy_c6} compares the fitted dimensions with the CFT states allowed at each dimension and lattice spin. Each fitted value represents a resolved eigenvalue branch; the listed states span the corresponding CFT subspace, not a one-to-one assignment of eigenvectors. The number of listed states can exceed the number of fitted branches because a finite lattice basis resolves only the independent combinations to which it overlaps. Noticeably, although all the resolved scaling dimensions from the eigenvalues of the operator covariance matrix are close to the theoretical predictions, their discrepancies are much larger than the fitting errors (over six standard deviations). This indicates the presence of the systematic errors: indeed, from Eq.~\ref{eq:main_result}, there are subleading corrections given by the mixing of the scaling operators and irrelevant perturbations.

\begin{table*}[t]
\caption{Point sets used to construct the odd-sector operators. The last column lists the lattice spins $\tilde{s}$ for which $P_{\tilde{s}}E(\mathbf x;X_{\mu})$ is nonzero.}
\label{tab:triangular_spin_lattice_operators}
\begin{ruledtabular}
\begin{tabular}{cll}
 & points in $X_{\mu}$ & allowed lattice spins $\tilde{s}$ \\
\hline
$X_1$ & $\{\mathbf0\}$ & $0$ \\
$X_2$ & $\{-\mathbf e_1,\mathbf0,\mathbf e_1\}$ & $0,2,4$ \\
$X_3$ & $\{\mathbf e_1,-\mathbf e_1+\mathbf e_2,-\mathbf e_2\}$ & $0,3$ \\
$X_4$ & $\{\mathbf0,\mathbf e_1,\mathbf e_2\}$ & $0,1,2,3,4,5$ \\
$X_5$ & $\{-2\mathbf e_1,\mathbf0,2\mathbf e_1\}$ & $0,2,4$ \\
$X_6$ & $\{2\mathbf e_1,-2\mathbf e_1+2\mathbf e_2,-2\mathbf e_2\}$ & $0,3$ \\
$X_7$ & $\{\mathbf0,\mathbf e_1,-2\mathbf e_1\}$ & $0,1,2,3,4,5$ \\
$X_8$ & $\{\mathbf e_1,-\mathbf e_1+\mathbf e_2,-2\mathbf e_2\}$ & $0,1,2,3,4,5$ \\
\end{tabular}
\end{ruledtabular}
\end{table*}

\begin{figure*}[t]
    \centering
    \includegraphics[width=\textwidth]{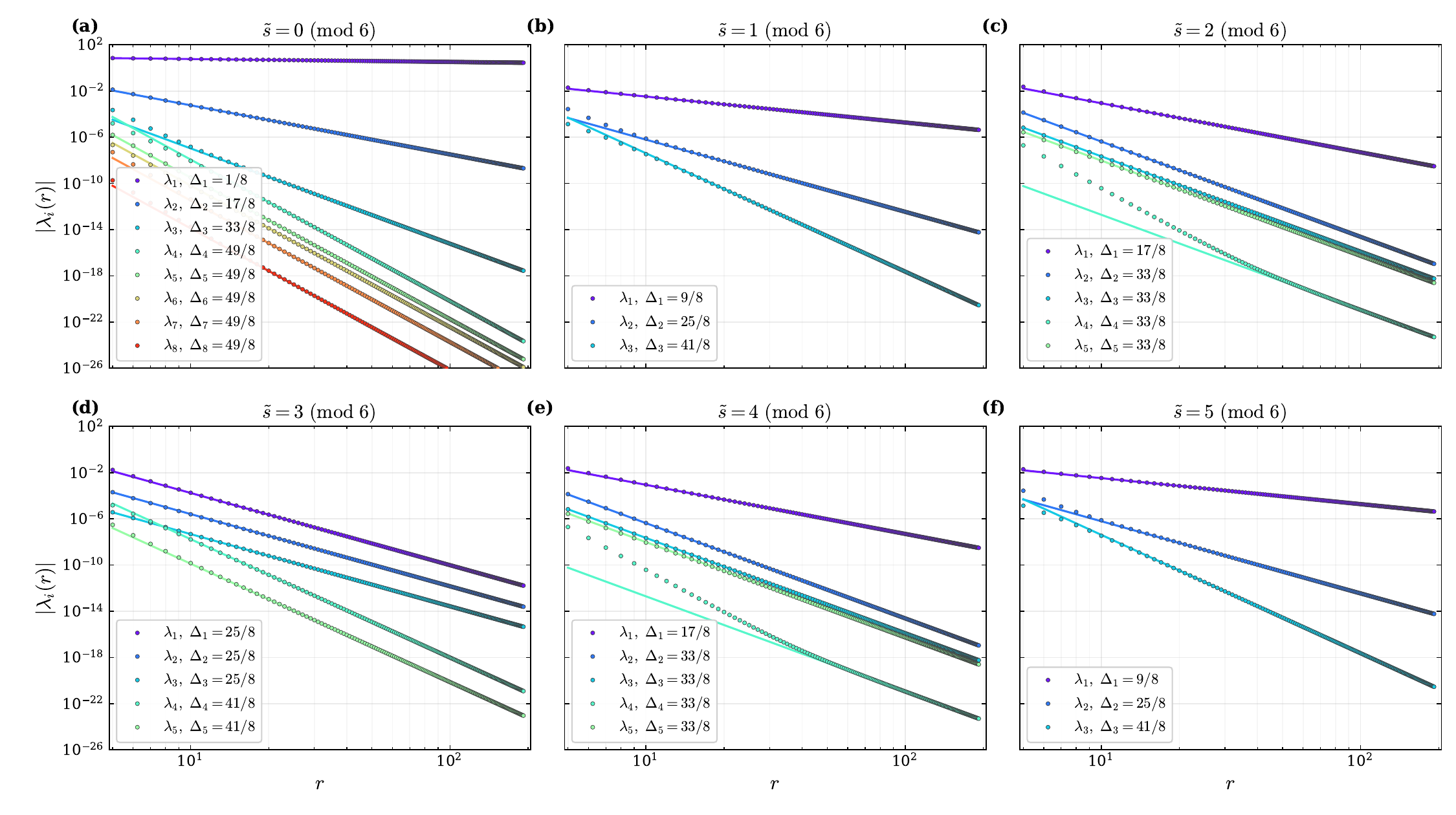}
    \caption{Odd-sector covariance spectra on the infinite triangular lattice for $5\leq r\leq192$. Panels (a)--(f) show $\tilde{s}=0,1,\ldots,5\pmod{6}$. The dots are the numerically exact plane data with the display threshold at $10^{-26}$. The legends identify the eigenvalue branches and their theoretical dimensions. The solid line associated with each branch has the theoretical slope $-2\Delta$ and passes through its last displayed points; these lines are not fits. The fitted scaling dimensions from the data over the range $96 \leq r \leq 192$ are listed in Table.~\ref{tab:triangular_spin_c6}.}
    \label{fig:triangular_spin_c6}
\end{figure*}

\subsubsection{Odd sector}
\label{sec:ising_spin_sector}

Odd products of spins have zero one-point functions by global spin-flip symmetry and couple to the $\sigma$ tower \cite{DiFrancesco1997}. We use eight point sets $X_1,\ldots,X_8$ in Tab.~\ref{tab:triangular_spin_lattice_operators} to construct the corresponding operators $P_{\tilde{s}}E(\mathbf x;X_{\mu})$. The six lattice-spin sectors have dimensions $(8,3,5,5,5,3)$ for $\tilde{s}=0,1,\ldots,5$, respectively, and account for 29 lattice operators. Figure~\ref{fig:triangular_spin_c6} shows the spectra, and Table~\ref{tab:triangular_spin_c6} gives the corresponding CFT assignments. As a representative example on how to interpret the results, we inspect the $\tilde{s}=0$ sector: it contains branches at $1/8$, $17/8$, and $33/8$, followed by five branches at $49/8$. The first corresponds to the spin primary, while the next two can be represented by scalar derivatives of $\sigma$ after using the null relations. The five branches at $49/8$ resolve several independent fields with the same dimension.

The multiplicities in the other sectors also contain information beyond derivative descendants. In the $\tilde{s}=2$ sector, four branches are resolved at $\Delta=33/8$. Derivatives of $\sigma$ alone would supply only two fields with this dimension and lattice spin: one with continuum spin $s=2$ and one with $s=-4$. The four resolved directions therefore include additional Virasoro descendants. The sectors $\tilde{s}=1$ and $\tilde{s}=5$ have coincident spectra, and so do the sectors $\tilde{s}=2$ and $\tilde{s}=4$ since they are related to each other by reflection.

Similar to the even sector, most incomplete multiplicities at larger scaling dimensions follow immediately from the size of the lattice basis. For example, only five directions remain at $\Delta=49/8$ in the scalar sector after its first three branches have been resolved, while the allowed CFT subspace is larger. However, there is one exception in the sector $(\Delta,\tilde{s})=(25/8,3)$: it contains four independent CFT states, but only three eigenvalue branches are resolved, even though the $\tilde{s}=3$ lattice basis has five operators. The total number of lattice operators therefore does not explain this missing branch.

To understand it, we inspect the reflection symmetry of the chosen lattice operators more carefully. Unlike the sectors $\tilde{s}=1,5$ and $\tilde{s}=2,4$ where reflection sends one lattice spin to its partner, for $\tilde{s}=3$, reflection keeps it invariant. As result, the $\tilde{s}=3$ sector can be divided into reflection-even and reflection-odd operators. On one hand, the four CFT states at $\Delta=25/8$ can be combined as
\begin{align}
    \Phi_{1,\pm}&=(L_{-3}\pm\bar L_{-3})\sigma,\notag\\
    \Phi_{2,\pm}&=(L_{-2}L_{-1}\pm
                       \bar L_{-2}\bar L_{-1})\sigma.
    \label{eq:ising_spin_parity_states}
\end{align}
Reflection exchanges the holomorphic and antiholomorphic expressions. Their sums are therefore even and their differences are odd, giving two CFT states of each parity. On the other hand, the chosen five-dimensional lattice basis, however, contains four reflection-even directions, constructed from $X_3,X_6,X_7,$ and $X_8$, and only one reflection-odd direction, constructed from $X_4$. The even operators cannot overlap with an odd CFT field, so their extra number cannot compensate for the shortage of odd operators. At $\Delta=25/8$, the maximum accessible multiplicity is consequently
\begin{equation}
    \min(4,2)+\min(1,2)=2+1=3.
    \label{eq:ising_spin_parity_count}
\end{equation}
The three observed branches exhaust this allowed multiplicity. The two remaining reflection-even lattice directions can still reveal higher-dimensional contributions and give the two branches at $\Delta=41/8$ in this sector. Thus, the missing fourth branch at $25/8$ is explained by the parity content of the lattice operator basis, rather than by the numerical precision or merely by an insufficient total number of lattice operators. Recovering it would require an additional independent reflection-odd lattice direction with a suitable overlap.

Table~\ref{tab:triangular_spin_c6} lists the CFT states after removal of null states and gives one fitted value for each resolved branch. Once again, although all the resolved scaling dimensions are extremely close to the theoretical predictions, their discrepancies are still far greater than the fitting errors due to the systematic errors from operator mixing and irrelevant perturbations.

\begin{table*}[t]
\caption{Odd-sector CFT states associated with the resolved eigenvalue branches. The lattice spin $\tilde{s}$ is understood modulo six. Fitted dimensions are obtained from the data over $96\leq r\leq192$ in Fig.~\ref{fig:triangular_spin_c6}. Fitting errors (as parenthesized) are formal regression standard errors.}
\label{tab:triangular_spin_c6}
\begin{ruledtabular}
\begin{tabular}{ccll}
$\tilde{s}\ (\mathrm{mod}\ 6)$ & $\Delta$ & CFT states & $\Delta_{\rm fit}$ \\
\hline
$0$ & $1/8$ & $\sigma$ & $0.1250147(3)$ \\
$0$ & $17/8$ & $L_{-1}\bar L_{-1}\sigma$ & $2.12499669(6)$ \\
$0$ & $33/8$ & $L_{-2}\bar L_{-2}\sigma$ & $4.125404(8)$ \\
$0$ & $49/8$ & $L_{-6}\sigma,\ L_{-5}L_{-1}\sigma,\ L_{-4}L_{-2}\sigma$ & $6.12436(1),\ 6.124695(6),\ 6.12430(1)$ \\
& & $L_{-3}^{2}\sigma,\ L_{-3}\bar L_{-3}\sigma,\ L_{-3}\bar L_{-2}\bar L_{-1}\sigma$ & $6.125064(2),\ 6.12575(1)$ \\
& & $L_{-2}L_{-1}\bar L_{-3}\sigma,\ L_{-2}L_{-1}\bar L_{-2}\bar L_{-1}\sigma,\ \bar L_{-6}\sigma$ & \\
& & $\bar L_{-5}\bar L_{-1}\sigma,\ \bar L_{-4}\bar L_{-2}\sigma,\ \bar L_{-3}^{2}\sigma$ & \\
$1$ & $9/8$ & $L_{-1}\sigma$ & $1.125132(2)$ \\
$1$ & $25/8$ & $L_{-2}\bar L_{-1}\sigma$ & $3.125363(7)$ \\
$1$ & $41/8$ & $L_{-3}\bar L_{-2}\sigma,\ L_{-2}L_{-1}\bar L_{-2}\sigma,\ \bar L_{-5}\sigma,\ \bar L_{-4}\bar L_{-1}\sigma,\ \bar L_{-3}\bar L_{-2}\sigma$ & $5.12446(1)$ \\
$2$ & $17/8$ & $L_{-2}\sigma$ & $2.125129(2)$ \\
$2$ & $33/8$ & $L_{-3}\bar L_{-1}\sigma,\ L_{-2}L_{-1}\bar L_{-1}\sigma,\ \bar L_{-4}\sigma,\ \bar L_{-3}\bar L_{-1}\sigma$ & $4.125073(1),\ 4.124815(3),\ 4.1415(5),\ 4.12447(1)$ \\
$3$ & $25/8$ & $L_{-3}\sigma,\ L_{-2}L_{-1}\sigma,\ \bar L_{-3}\sigma,\ \bar L_{-2}\bar L_{-1}\sigma$ & $3.125171(3),\ 3.124613(7),\ 3.1249530(9)$ \\
$3$ & $41/8$ & $L_{-4}\bar L_{-1}\sigma,\ L_{-3}L_{-1}\bar L_{-1}\sigma,\ L_{-1}\bar L_{-4}\sigma,\ L_{-1}\bar L_{-3}\bar L_{-1}\sigma$ & $5.124790(4),\ 5.12558(1)$ \\
$4$ & $17/8$ & $\bar L_{-2}\sigma$ & $2.125129(2)$ \\
$4$ & $33/8$ & $L_{-4}\sigma,\ L_{-3}L_{-1}\sigma,\ L_{-1}\bar L_{-3}\sigma,\ L_{-1}\bar L_{-2}\bar L_{-1}\sigma$ & $4.125073(1),\ 4.124815(3),\ 4.1415(5),\ 4.12447(1)$ \\
$5$ & $9/8$ & $\bar L_{-1}\sigma$ & $1.125132(2)$ \\
$5$ & $25/8$ & $L_{-1}\bar L_{-2}\sigma$ & $3.125363(7)$ \\
$5$ & $41/8$ & $L_{-5}\sigma,\ L_{-4}L_{-1}\sigma,\ L_{-3}L_{-2}\sigma,\ L_{-2}\bar L_{-3}\sigma,\ L_{-2}\bar L_{-2}\bar L_{-1}\sigma$ & $5.12446(1)$ \\
\end{tabular}
\end{ruledtabular}
\end{table*}

\subsubsection{Extraction of $k_{ab}$ for the triangular lattice}
\label{sec:ising_eigenvector_test}

We have shown that the eigenvalue spectra identify scaling dimensions and accessible multiplicities. We now turn to our second objective: extracting two-point amplitudes from the covariance eigenvectors. We restrict this test to the lattice-spin $\tilde{s}=0$ sectors in the even and odd sectors. At each construction distance $R$, we diagonalize the covariance matrix, use its eigenvectors to form the orthogonal lattice operators in Eq.~\eqref{eq:eigenoperator}, and measure their normalized cross ratios $\rho_{ab}(r;R)$ in Eq.~\eqref{eq:rho_definition} at separations $r$. Notice that the orthogonal lattice operators constructed at finite $R$ are not themselves scaling fields, as explained in Sec.~\ref{sec:covariance_eigenvectors_theory}.

\begin{figure*}[t]
    \centering
    \includegraphics[width=\textwidth]{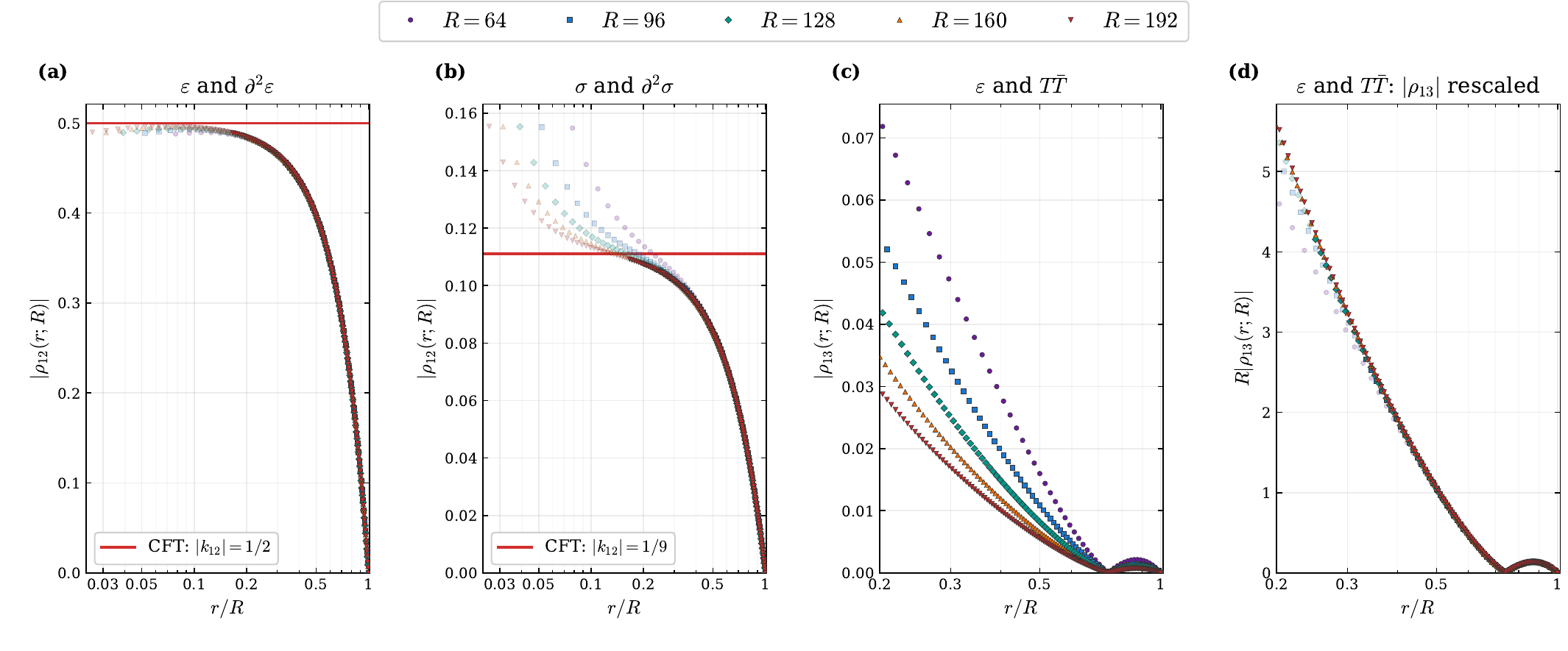}
    \caption{
        Finite-distance tests of the covariance eigenvectors in the $\tilde{s}=0$ blocks on the infinite triangular lattice, plotted against $r/R$ for $R=64,96,128,160,$ and $192$. Panels (a) and (b) show $|\rho_{12}(r;R)|$ for $(\varepsilon,\partial^2\varepsilon)$ in the even sector and $(\sigma,\partial^2\sigma)$ in the odd sector, respectively, with $5\leq r\leq R$. The horizontal red lines mark the plane-CFT two-point amplitudes $|k_{12}|=1/2$ and $1/9$. Panels (c) and (d) show the cross-tower correlation between $\varepsilon$ and $T\bar T$: panel (c) displays the unrescaled magnitude $|\rho_{13}(r;R)|$, while panel (d) displays $R|\rho_{13}(r;R)|$. These two panels use $r/R\geq0.2$. Every curve vanishes at $r/R=1$ by construction. Points with $r<32$ are shaded in panels (a), (b), and (d) to indicate the short-distance region but are not excluded; the unrescaled data in panel (c) are shown without shading.
    }
    \label{fig:ising_eigenvector_test}
\end{figure*}

We first study the magnitude of the normalized cross ratio $|\rho_{12}(r;R)|$ between the leading two scalar branches, and as predicted in Eq.~\ref{eq:rho_scaling_limit}, we anticipate
\begin{equation}
    \lim_{r \rightarrow +\infty}\left[\lim_{R \rightarrow +\infty}|\rho_{12}(r;R)|\right] = |k_{12}|
    \label{eq:triangular_lattice_rho12}
\end{equation}
In the lattice-spin $\tilde{s}=0$ sectors, $|k_{12}|$ is determined by the primary $\phi$ and its Laplacian descendant $\partial^2\phi$: these are $(\varepsilon,\partial^2\varepsilon)$ in the even sector and $(\sigma,\partial^2\sigma)$ in the odd sector. Thus, the theoretical value of $|k_{12}|$ can be calculated explicitly: starting with $\langle\phi(0)\phi(\mathbf r)\rangle= r^{-2\Delta_\phi}$, where we have set the normalization $k_{11}=1$, differentiation in two dimensions gives
\begin{align}
    &\langle\phi(0)\partial^2\phi(\mathbf r)\rangle
    =4\Delta_\phi^2 r^{-2\Delta_\phi-2},\notag\\
    &\langle\partial^2\phi(0)\partial^2\phi(\mathbf r)\rangle
    =16\Delta_\phi^2(\Delta_\phi+1)^2
      r^{-2\Delta_\phi-4}.
    \label{eq:ising_laplacian_correlators}
\end{align}
The normalization and the separation dependence cancel in the ratio, leaving
\begin{equation}
    |k_{12}|
    =\frac{\Delta_\phi}{\Delta_\phi+1}.
    \label{eq:ising_rho_cft}
\end{equation}
The predicted amplitudes are therefore $|k_{12}|=1/2$ in the even sector and $|k_{12}|=1/9$ in the odd sector.

To demonstrate how $|k_{12}|$ can be extracted from the normalized cross ratio $|\rho_{12}(r;R)|$, as explained in Sec.~\ref{sec:covariance_eigenvectors_theory}, we perform a scaling-collapse analysis by plotting $|\rho_{12}(r;R)|$ against $r/R$. In the scale-separated regime $0<r/R\ll 1$, we can inspect the collapsed curve and study how it approaches the theoretical value. Panels (a) and (b) of Fig.~\ref{fig:ising_eigenvector_test} show $|\rho_{12}(r;R)|$ for $R=64,96,128,160,$ and $192$. In both the even and odd sectors, the curves approximately collapse over intermediate $r/R$ and agree well with the corresponding CFT amplitudes. Points with $r<32$ are shaded to distinguish the short-distance region but remain in the plot.

As the next nontrivial test of the prediction in Eq.~\ref{eq:rho_scaling_limit}, we study the orthogonality of fields from different conformal towers. The $\tilde{s}=0$ sector of the even sector provides such a test. Its first and third branches approach $\varepsilon$ and $T\bar T=L_{-2}\bar L_{-2}\mathbb I$, respectively. Since these fields belong to different conformal towers, their plane two-point amplitude vanishes: $k_{13}=0$ \cite{DiFrancesco1997}. Through measuring $|\rho_{13}(r;R)|$, we can demonstrate the orthogonality between the fields across the conformal towers as a strong check of the CFT.

 Hence, as the first step, we examine the magnitude $|\rho_{13}(r;R)|$ and study how it varies with $r/R$ in panel (c) of Fig.~\ref{fig:ising_eigenvector_test}. However, since the leading continuum contribution to $\mathcal C_{13}(r;R)$ vanishes, the direct plot of $|\rho_{13}(r;R)|$ against $r/R$ does not collapse for different choices of $R$. To treat this, we look into the next scalar field in the $\varepsilon$ tower above dimension four is $L_{-2}\bar L_{-2}\varepsilon$ at $\Delta=5$, which can correlate with $\varepsilon$. At fixed $r/R$, this channel contributes at order $R^{-(1+5)}=R^{-6}$, while the normalization in $\rho_{13}$ is of order $R^{-(1+4)}=R^{-5}$. Orthogonality between the towers removes the otherwise possible $R^{-5}$ term; the lower-dimensional admixtures induced by diagonalization enter at the same $R^{-6}$ order and determine the $r/R$ dependence, including the zero at $r/R=1$. Hence, the first allowed correction has the form $|\rho_{13}(r;R)|=R^{-1}|F^{(1)}_{13}(r/R)|+\ldots$, provided its coefficient does not vanish. Consequently, in panel (d), the curves of $R|\rho_{13}(r;R)|$ display a good approximate collapse. This behavior is consistent with a vanishing continuum two-point amplitude and a nonzero $R^{-1}$ finite-distance correction. This therefore illustrates that how orthogonality between different conformal towers can be examined with the operator covariance method.

\section{Discussion}

Extracting quantitative information about the underlying continuum theory from studies of lattice statistical mechanics model at its critical point is a challenging problem. An interesting approach proposed in Ref.~\cite{Sandvik2026}, is to compute the covariance matrix constructed from two points of lattice operators and study its eigenspectrum. It was found that the eigenvalues leading decay with distance as power laws with exponents equal to the smallest scaling dimensions. While this result was tested for various models in Ref.~\cite{Sandvik2026}, a proof for its validity was not provided.  Here we have provided an explicit proof of the eigenvalue spectrum leading behavior simply from the scaling ansatz (we do not assume conformal invariance). Based on insights from our proof we have proposed a new method to extract the two-point amplitudes, another universal piece of data, from the eigenvectors of the covariance matrix. Taken together the eigen-decomposition of the operator covariance matrix can deduce the full universal structure of the two-point correlation functions, leading to various new tests of critical phenomena, including the orthogonality of scaling operators that belong to distinct conformal towers. All of our results have been tested thoroughly for synthetic data and in detail for the triangular and square Ising models, where thanks to exact methods we can study a very large number of scaling operators.  

While the diagonalization of the covariance matrix is clearly very powerful and can have wide applications, we comment here on some open problems that would be interesting to solve in future developments of this method. It is not yet known how to extract the expansion coefficients in Eq.~\eqref{eq:lattice_expansion}. These expansion coefficients are well defined for a model, but are non-universal model dependent quantities. Even so determining these numbers could help in further analysis, for instance, to create accurate lattice realizations of scaling operators  through systematically canceling subleading terms. This can be used to study more accurate conformal data, like the operator product expansion coefficients, as was done in (1+1)D CFT~\cite{vidal2020}. Another weakness is that the eigenvalues suffer from large sub-leading corrections, which even do not improve the accuracy over fitting the two point correlation function of a single operator. In other words, while the diagonalization improves our ability to resolve more scaling operators conveniently, it does not improve the accuracy with which the leading scaling dimension can be measured. Finally, it would be interesting to understand how this method can be used to study three-point correlation functions, so more CFT data can be extracted.

\begin{acknowledgments}
The authors acknowledge financial support from NSF DMR-2312742. Y.-C.C. also acknowledges financial support from the Ministry of Education, Taiwan, through the 2026 Study Abroad Scholarship. This work used the Anvil supercomputer at Purdue University through allocation PHY260077 from the Advanced Cyberinfrastructure Coordination Ecosystem: Services \& Support (ACCESS) program, which is supported by the U.S. National Science Foundation under Grant Nos. 2138259, 2138286, 2138307, 2137603, and 2138296~\cite{Boerner2023}. Anvil is supported by the National Science Foundation under Grant No. 2005632~\cite{Song2022}. We also acknowledge the Penn State Institute for Computational and Data Sciences (ICDS; RRID:SCR\_025154) for providing access to the Roar Collab computational resources (RRID:SCR\_026424).
\end{acknowledgments}

\appendix

\section{Direct Determinant Proof}
\label{app:determinant}

Here we prove the eigenvalue asymptotics in Eq.~\eqref{eq:main_result} directly from the characteristic determinant. To start, recall the $N\times N$ Hermitian graded matrix defined in Eq.~\eqref{eq:M_limit}
\begin{equation}
    G_{ij}(r)=\frac{M_{ij}(r)}{r^{\Omega_i+\Omega_j}},
    \qquad
    \lim_{r\rightarrow\infty}M_{ij}(r)=m_{ij},
    \label{eq:graded_matrix_app}
\end{equation}
where $\Omega_1\leq\Omega_2\leq\cdots\leq\Omega_N$. We let
\begin{equation}
    \Omega^{(1)}<\Omega^{(2)}<\cdots<\Omega^{(n)}
    \label{eq:distinct_dimensions_app}
\end{equation}
be the distinct exponents among $\Omega_1,\ldots,\Omega_N$, and define
\begin{equation}
    I_p=\{i\in\{1,\ldots,N\}:\Omega_i=\Omega^{(p)}\},
    \qquad
    d_p=|I_p|.
    \label{eq:scaling_blocks_app}
\end{equation}
Thus, $I_p$ contains all indices with exponent $\Omega^{(p)}$, $d_p$ is its degeneracy, and $\sum_{p=1}^{n}d_p=N$.

As will be useful later, we take a pause to carefully denote the submatrices of the limiting matrix $m$ that appear below. For each $p$, let
\begin{equation}
    m_p=(m_{ij})_{i,j\in I_p}
    \label{eq:current_coefficient_block_app}
\end{equation}
be the $d_p\times d_p$ coefficient block associated with $\Omega^{(p)}$. We use $<p$ to denote the union $I_1\cup\cdots\cup I_{p-1}$ of all blocks with smaller exponents and define
\begin{equation}
    m_{<p}=(m_{ij})_{i,j\in I_1\cup\cdots\cup I_{p-1}}.
    \label{eq:lower_coefficient_block_app}
\end{equation}
The matrix connecting these blocks with smaller exponents to block $p$ is
\begin{equation}
    b_p=(m_{ij})_{i\in I_1\cup\cdots\cup I_{p-1},\,j\in I_p}.
    \label{eq:coefficient_mixing_block_app}
\end{equation}
Since $m$ is Hermitian, the block in the opposite direction is $b_p^{\dagger}$. The cumulative coefficient matrix containing all blocks through $p$ is therefore
\begin{equation}
    m_{\leq p}
    =
    \begin{pmatrix}
        m_{<p}&b_p\\
        b_p^{\dagger}&m_p
    \end{pmatrix}.
    \label{eq:cumulative_coefficient_block_app}
\end{equation}
For $p=1$, $m_{<1}$ and $b_1$ are absent, and $m_{\leq1}=m_1$. For $p=n$, we recover $m=m_{\leq n}$.

Our objective is to find the eigenvalues of $G(r)$, which are the roots of
\begin{equation}
    \det\!\left[G(r)-\lambda I\right]=0.
    \label{eq:characteristic_equation_app}
\end{equation}
Given Eq.~\eqref{eq:characteristic_equation_app}, we factor $r^{-\Omega_j}$ from column $j$ and $r^{-\Omega_i}$ from row $i$ and have
\begin{equation}
    \det\!\left[G(r)-\lambda I\right]
    =r^{-2\sum_{i=1}^{N}\Omega_i}
    \det\!\left[M(r)-\lambda Q(r)\right],
    \label{eq:factored_characteristic_app}
\end{equation}
where $Q(r)$ is a diagonal matrix with $Q_{ij}(r)=r^{2\Omega_i}\delta_{ij}$. Since the prefactor $r^{-2\sum_{i=1}^{N}\Omega_i}$ is nonzero, the characteristic equation is equivalent to
\begin{equation}
    \det\!\left[M(r)-\lambda Q(r)\right]=0.
    \label{eq:reduced_characteristic_app}
\end{equation}
This form separates the finite coefficient matrix $M(r)$ from all explicit powers of $r$ multiplying $\lambda$.

The strategy to prove the asymptotic behavior of the eigenvalues is actually quite simple: through the direct construction, we derive the eigenvalues associated with the exponent $\Omega^{(p)}$. As a reasonable ansatz, we start with
\begin{equation}
    \lambda(r)
    =r^{-2\Omega^{(p)}}[c+\epsilon(r)],
    \label{eq:eigenvalue_ansatz_app}
\end{equation}
where the constant $c$ will be determined by the leading characteristic equation. We also assume 
\begin{equation}
    \lim_{r \rightarrow +\infty}\epsilon(r)=0,
\end{equation}
in our ansatz, which will be proved explicitly once we calculate the solution of the characteristic equation. Note that
\begin{equation}
    \lambda(r)Q_{ii}(r)=[c+\epsilon(r)]r^{2(\Omega_i-\Omega^{(p)})}.
    \label{eq:scaled_diagonal_term_app}
\end{equation}
Hence, in the large-$r$ limit, the $i$-th diagonal component of $M(r)-\lambda Q(r)$ follows
\begin{equation}
    M_{ii}(r)-\lambda(r)Q_{ii}(r)
    \longrightarrow
    \begin{cases}
        m_{ii},&\Omega_i<\Omega^{(p)},\\
        m_{ii}-c,&\Omega_i=\Omega^{(p)},\\
        \infty,&\Omega_i>\Omega^{(p)}.
    \end{cases}
    \label{eq:three_block_limits_app}
\end{equation}
Thus, the $\lambda$-dependent term vanishes in every block below $p$, remains finite in block $p$, and dominates the finite entries of $M(r)$ in every block above $p$.

Hence, in the determinant expansion, the leading contribution from every index in a block above $p$ must contain its large diagonal factor, and after factoring out the dominant diagonal contributions from all blocks above $p$, Eq.~\eqref{eq:reduced_characteristic_app} becomes
\begin{align}
    0={}&
    \prod_{q>p}
    \left\{-[c+\epsilon(r)]r^{2(\Omega^{(q)}-\Omega^{(p)})}\right\}^{d_q}
    \notag\\
    &\times
    \left[
    \det
    \begin{pmatrix}
        m_{<p}&b_p\\
        b_p^{\dagger}&m_p-[c+\epsilon(r)]I_{d_p}
    \end{pmatrix}
    +\Theta(r)
    \right],
    \label{eq:leading_determinant_expansion_app}
\end{align}
where
\begin{equation}
    \lim_{r\rightarrow +\infty}\Theta(r) =0.
    \label{eq:Theta_magnitude}
\end{equation}
The equation for $c$ to fulfill Eq.~\eqref{eq:leading_determinant_expansion_app} is therefore
\begin{equation}
    \det
    \begin{pmatrix}
        m_{<p}&b_p\\
        b_p^{\dagger}&m_p-cI_{d_p}
    \end{pmatrix}
    =0.
    \label{eq:leading_coefficient_equation_app}
\end{equation}
When $p=1$, the block is just $m_1-cI_{d_1}$, and we can derive all $d_1$ roots through diagonalization.

For $p>1$, provided that $m_{<p}$ is non-singular, we apply the Schur-complement determinant identity to Eq.~\eqref{eq:leading_coefficient_equation_app} and obtain \cite{HornJohnson2013,Zhang2005}
\begin{equation}
    0=\det m_{<p} \times \det\!\left[
        m_p-b_p^{\dagger}m_{<p}^{-1}b_p-cI_{d_p}
    \right].
    \label{eq:block_determinant_coefficient_app}
\end{equation}
We define the effective coefficient matrix for the $p$-th block by
\begin{equation}
    \widehat m_p=m_p-b_p^{\dagger}m_{<p}^{-1}b_p.
    \label{eq:effective_coefficient_block_app}
\end{equation}
The second term contains the effect of mixing with every lower-dimensional block. Since $m$ is Hermitian, $\widehat m_p$ is Hermitian as well. Let its eigenvalues be
\begin{equation}
    \kappa^{(p)}_1,\ldots,\kappa^{(p)}_{d_p}.
\end{equation}
The solutions of the leading equation are then $c=\kappa^{(p)}_\rho$, with $\rho=1,\ldots,d_p$.

Now we only have to show $\lim_{r \rightarrow +\infty}\epsilon(r)=0$. This is evident from Eq.~\eqref{eq:Theta_magnitude}: since $\lim_{r\rightarrow +\infty}\Theta(r)=0$, there exists $\epsilon(r)$ which also vanishes in the large-$r$ limit with $c+\epsilon(r)$ still satisfying Eq.~\eqref{eq:leading_determinant_expansion_app}. Hence, combining all solutions of $\lambda(r)$ for the $p$-th block, we have constructed its all eigenvalues:
\begin{equation}
    \lambda^{(p)}_\rho(r)
    =r^{-2\Omega^{(p)}}
    \left[\kappa^{(p)}_\rho+\epsilon^{(p)}_{\rho}(r)\right],
    \qquad
    \rho=1,\ldots,d_p.
    \label{eq:degenerate_result_app}
\end{equation}
where $\epsilon^{(p)}_{\rho}(r)$ vanishes in the large-$r$ limit. Thus,
\begin{equation}
    \lambda^{(p)}_\rho(r)
    \sim\kappa^{(p)}_\rho r^{-2\Omega^{(p)}}.
    \label{eq:degenerate_asymptotic_result_app}
\end{equation}
Therefore, the block with exponent $\Omega^{(p)}$ produces $d_p$ eigenvalues with the common decay exponent $2\Omega^{(p)}$. Repeating the construction for every block $p=1,\ldots,n$ produces $\sum_{p=1}^{n}d_p=N$ roots and accounts for the complete spectrum of $G(r)$.

In the previous proof, we have assumed that $\kappa^{(p)}_\rho\neq0$. We now discuss what condition would guarantee Eq.~\eqref{eq:degenerate_asymptotic_result_app}. We start with the case when all exponents are non-degenerate (i.e. $d_p=1$ for every $p$). The effective matrix $\widehat m_p$ in Eq.~\eqref{eq:effective_coefficient_block_app} is then a scalar. From Eqs.~\eqref{eq:cumulative_coefficient_block_app} and \eqref{eq:effective_coefficient_block_app},
\begin{equation}
    \det m_{\leq p}=\det m_{<p}\,\kappa^{(p)},
\end{equation}
and hence
\begin{equation}
    \kappa^{(p)}=\frac{\det m_{\leq p}}{\det m_{<p}}.
    \label{eq:nondegenerate_amplitude_app}
\end{equation}
The corresponding eigenvalue is
\begin{equation}
    \lambda^{(p)}(r)
    =r^{-2\Omega^{(p)}}
    \left[
    \frac{\det m_{\leq p}}{\det m_{<p}}
    +\epsilon^{(p)}(r)
    \right],
    \label{eq:nondegenerate_result_app}
\end{equation}
where the denominator for $p=1$ equals one and $\epsilon^{(p)}(r)$ vanishes as $r\rightarrow\infty$.

We then turn to the general case with the non-vanishing effective-block condition, which ensures that all the eigenvalues have the asymptotic behavior in Eq.~\ref{eq:nondegenerate_result_app}. This condition has two criteria:
\begin{equation}
    \det m_{<p}\neq0\quad(p>1),
    \qquad
    \det\widehat m_p\neq0\quad(p=1,\ldots,n).
    \label{eq:visibility_condition_app}
\end{equation}
The second criterion guarantees that none of the coefficients $\kappa^{(p)}_\rho$ vanishes. If one of these coefficients is zero, the corresponding eigenvalue has no nonzero leading term proportional to $r^{-2\Omega^{(p)}}$, and its first non-vanishing contribution must be determined from the subleading terms. Positive definiteness of $m$ is sufficient, but not necessary, for these non-singularity conditions.

As a final remark, the $r$ dependence of $M(r)$ changes only the functions $\epsilon^{(p)}_\rho(r)$. In particular, the irrelevant correction in Eq.~\eqref{eq:K_asymptotic} modifies the approach to the asymptotic regime but not the exponents in Eq.~\eqref{eq:degenerate_asymptotic_result_app}.

\section{Covariance Eigenvectors and Two-Point Amplitudes}

\subsection{Asymptotic Form of the Orthogonal Lattice Operators}
\label{app:eigenoperator_asymptotics}

Here we derive the coefficient behavior in Eq.~\eqref{eq:eigenoperator_in_scaling_operator_expression} directly from the covariance eigenvector equation and the eigenvalue result of Appendix~\ref{app:determinant}. This derivation uses only the scaling form of the covariance matrix and does not assume conformal invariance at criticality. We first express the eigenvector equation in terms of the overlap coefficients $q_{ai}(R)$. We then analyze separately the coefficients below, above, and within the scaling-dimension block associated with the eigenvalue branch. We initially assume that the retained fields end at a scaling-dimension gap, $\Delta_{N_\ell}<\Delta_{N_\ell+1}$, so that $\overline K(R)$ approaches the corresponding block of $k$. The case in which the cutoff lies inside a degenerate block is addressed at the end.

Because $\ket{w_1},\ldots,\ket{w_{N_\ell}}$ are linearly independent, their Gram matrix
\begin{equation}
    \mathcal G_{ab}=\braket{w_a|w_b},
    \qquad 1\leq a,b\leq N_\ell,
    \label{eq:eigenoperator_gram_app}
\end{equation}
is positive definite. The matrix $\mathcal G$ accounts for the fact that the overlap vectors need not be orthogonal. For a normalized covariance eigenvector $\ket{u_i(R)}$, let $q_{ai}(R)=\braket{w_a|u_i(R)}$, as in the main text. Expanding the eigenvector in this basis gives
\begin{equation}
        \ket{u_i(R)}
        =\sum_{a,b=1}^{N_\ell}\ket{w_a}
        (\mathcal G^{-1})_{ab}q_{bi}(R)
    \label{eq:eigenoperator_gram_normalization_app}
\end{equation}
where $q_i^{\dagger}(R)\mathcal G^{-1}q_i(R)=1$. In particular, each $q_{ai}(R)$ is bounded. Substituting Eq.~\eqref{eq:finite_covariance} into $C(R)\ket{u_i(R)}=\lambda_i(R)\ket{u_i(R)}$ and equating the coefficients of the independent vectors $\ket{w_a}$ yields the exact equation
\begin{equation}
    R^{-\Delta_a}\sum_{b=1}^{N_\ell}
    \overline K_{ab}(R)R^{-\Delta_b}q_{bi}(R)
    =\lambda_i(R)\sum_{b=1}^{N_\ell}
    (\mathcal G^{-1})_{ab}q_{bi}(R).
    \label{eq:eigenoperator_exact_equation_app}
\end{equation}

Equation~\eqref{eq:eigenoperator_exact_equation_app} is the starting point of the proof. Consider an eigenvalue branch $i$ associated with the block of scaling dimension $\Delta^{(p)}$. Appendix~\ref{app:determinant} establishes that
\begin{equation}
    \lim_{R\rightarrow\infty}
    R^{2\Delta^{(p)}}\lambda_i(R)
    =\kappa^{(p)}_{\rho}\neq0,
    \label{eq:eigenoperator_eigenvalue_limit_app}
\end{equation}
where $\rho$ labels the branch within that block. We use $<p$, $p$, and $>p$ to denote the retained indices with scaling dimensions below, equal to, and above $\Delta^{(p)}$, respectively.

We begin with the coefficients below the target block. For $a\in<p$, define
\begin{equation}
    A_{ai}(R)=R^{\Delta^{(p)}-\Delta_a}q_{ai}(R).
    \label{eq:eigenoperator_rescaled_lower_app}
\end{equation}
The required suppression of $q_{ai}(R)$ will follow once we show that $A_{ai}(R)$ remains bounded; it is not assumed in this definition.

For $a\in<p$, divide Eq.~\eqref{eq:eigenoperator_exact_equation_app} by $R^{-\Delta_a-\Delta^{(p)}}$ to obtain
\begin{widetext}
\begin{equation}
        \sum_{b\in<p}\overline K_{ab}(R)A_{bi}(R)
        +\sum_{b\in p}\overline K_{ab}(R)q_{bi}(R)
        +\sum_{b\in>p}\overline K_{ab}(R)
        R^{-(\Delta_b-\Delta^{(p)})}q_{bi}(R)
        =\bigl[R^{2\Delta^{(p)}}\lambda_i(R)\bigr]
        R^{-(\Delta^{(p)}-\Delta_a)}(\mathcal G^{-1}q_i(R))_a.
    \label{eq:eigenoperator_rescaled_equation_app}
\end{equation}
\end{widetext}
The third term on the left-hand side and the right-hand side vanish at large $R$, because their explicit powers of $R$ decay and the coefficients $q_{ai}(R)$ are bounded. The remaining terms therefore satisfy
\begin{equation}
    \overline K_{<p,<p}(R)A_{<p,i}(R)
    +\overline K_{<p,p}(R)q_{p,i}(R)
    \longrightarrow0.
    \label{eq:eigenoperator_lower_equation_app}
\end{equation}
The limiting lower principal block $k_{<p,<p}$ is non-singular under the non-vanishing effective-block condition of Appendix~\ref{app:determinant}: the Gram--Schmidt transformation restricted to the lower blocks is invertible and preserves their non-singularity. We can therefore solve Eq.~\eqref{eq:eigenoperator_lower_equation_app} for $A_{<p,i}(R)$. This establishes its boundedness and gives the sharper relation
\begin{equation}
    A_{<p,i}(R)
    +k_{<p,<p}^{-1}k_{<p,p}q_{p,i}(R)
    \longrightarrow0.
    \label{eq:eigenoperator_lower_limit_app}
\end{equation}
Thus, every coefficient with $\Delta_a<\Delta^{(p)}$ contains the factor $R^{-(\Delta^{(p)}-\Delta_a)}$ appearing in Eq.~\eqref{eq:eigenoperator_in_scaling_operator_expression}, and $q_{<p,i}(R)\rightarrow0$.

We next consider the coefficients above the target block. For $a\in>p$, after using Eq.~\eqref{eq:eigenoperator_lower_limit_app}, the left-hand side of Eq.~\eqref{eq:eigenoperator_exact_equation_app} is $R^{-\Delta_a-\Delta^{(p)}}$ times a bounded expression. Since the rescaled eigenvalue has the nonzero limit in Eq.~\eqref{eq:eigenoperator_eigenvalue_limit_app}, we obtain $(\mathcal G^{-1}q_i(R))_{>p}\rightarrow0$. The $>p,>p$ principal block of $\mathcal G^{-1}$ is positive definite, so this implies
\begin{equation}
    q_{>p,i}(R)
    +(\mathcal G^{-1})_{>p,>p}^{-1}
    (\mathcal G^{-1})_{>p,p}q_{p,i}(R)
    \longrightarrow0.
    \label{eq:eigenoperator_higher_limit_app}
\end{equation}
If there are no indices above $p$, this relation is simply omitted. This gives how coefficients above the target block are related to those in the target block.

It remains to determine the component within the target block. For this purpose, define the two finite matrices
\begin{align}
    S_p={}&k_{p,p}
    -k_{p,<p}k_{<p,<p}^{-1}k_{<p,p},
    \notag\\
    \Gamma_p={}&(\mathcal G^{-1})_{p,p}
    -(\mathcal G^{-1})_{p,>p}
    (\mathcal G^{-1})_{>p,>p}^{-1}
    (\mathcal G^{-1})_{>p,p}.
    \label{eq:eigenoperator_effective_blocks_app}
\end{align}
Terms involving an empty $<p$ or $>p$ set are absent. The matrix $S_p$ incorporates the cancellation produced by the lower-dimensional coefficients, while $\Gamma_p$ gives the normalization induced by the non-orthogonal overlap vectors after the higher-dimensional coefficients are eliminated. In particular, $\Gamma_p$ is positive definite. Rescaling the $p$ rows of Eq.~\eqref{eq:eigenoperator_exact_equation_app} and using Eqs.~\eqref{eq:eigenoperator_lower_limit_app} and \eqref{eq:eigenoperator_higher_limit_app} gives
\begin{equation}
    S_pq_{p,i}(R)
    -\bigl[R^{2\Delta^{(p)}}\lambda_i(R)\bigr]
    \Gamma_pq_{p,i}(R)
    \longrightarrow0.
    \label{eq:eigenoperator_target_equation_app}
\end{equation}
Equations~\eqref{eq:eigenoperator_gram_normalization_app}, \eqref{eq:eigenoperator_lower_limit_app}, and \eqref{eq:eigenoperator_higher_limit_app} also imply
\begin{equation}
    q_{p,i}^{\dagger}(R)\Gamma_pq_{p,i}(R)
    \longrightarrow1.
    \label{eq:eigenoperator_target_normalization_app}
\end{equation}
The target-block vector therefore cannot vanish. Every one of its limiting vectors $v_i$ satisfies the finite generalized eigenvalue equation
\begin{equation}
    S_pv_i=\kappa^{(p)}_{\rho}\Gamma_pv_i,
    \qquad v_i^{\dagger}\Gamma_pv_i=1.
    \label{eq:eigenoperator_target_limit_app}
\end{equation}

When $\kappa^{(p)}_{\rho}$ is simple, Eq.~\eqref{eq:eigenoperator_target_limit_app} fixes $v_i$ up to the phase of $\ket{u_i(R)}$. Choosing that phase consistently makes $q_{p,i}(R)$ converge to $v_i$. Equations~\eqref{eq:eigenoperator_lower_limit_app} and \eqref{eq:eigenoperator_higher_limit_app} then give limits for the rescaled lower-dimensional coefficients and the higher retained coefficients. For a nondegenerate scaling dimension, the target block contains a single field, and its coefficient has a nonzero limit. The eigenvector itself also converges in the finite lattice-operator space, so $q_{ai}(R)=\braket{w_a|u_i(R)}$ has a limit for each fixed field beyond the cutoff.

Combining the three parts of the argument gives Eq.~\eqref{eq:eigenoperator_in_scaling_operator_expression}: coefficients below the target dimension are suppressed by $R^{-(\Delta^{(p)}-\Delta_a)}$, the target-block coefficient remains nonzero, and coefficients above the target dimension remain bounded. The infinite scaling-field expansion is understood with the same tail-convergence assumption used for the covariance matrix.

When $\Delta^{(p)}$ is degenerate, the same argument applies to the entire target block. If several branches also share the same leading coefficient $\kappa^{(p)}_{\rho}$, Eq.~\eqref{eq:eigenoperator_target_limit_app} determines their limiting subspace but need not select a unique limit for each eigenvector, because subleading terms may rotate the vectors within that subspace. The lower-dimensional suppression and boundedness of all coefficients still follow, while individual coefficient limits require a choice of basis or additional control of the subleading terms. If the cutoff lies inside a degenerate multiplet, the proof instead uses $\lim_{R\rightarrow\infty}\overline K(R)$ in place of $k$ in the block equations. This limit retains the contributions from the omitted fields with the same scaling dimension.

After showing Eq.~\eqref{eq:eigenoperator} explicitly, we now proceed to establish the scaling form in Eq.~\eqref{eq:scaling_form_of_projected_covariance}. To begin, we substitute Eq.~\eqref{eq:eigenoperator_in_scaling_operator_expression} into Eq.~\eqref{eq:projected_covariance} and have
\begin{equation}
    \left\langle\widetilde O_a^\dagger(0;R)\widetilde O_b(r;R)\right\rangle = \sum_{c,d=1}^{+\infty}q^*_{ca}(R)q_{db}(R)\langle \phi^{\dagger}_c(0)\phi_d(r) \rangle.
    \label{eq:orthogonal_lattice_operator_correlations_classification}
\end{equation}
We classify the resulting terms according to their orders of magnitude in $r$ and $R$. There are three classes of contributions. First, given $\Delta_c<\Delta_a$ and $\Delta_d<\Delta_b$, the contribution from $\langle\phi_c^\dagger(0)\phi_d(r)\rangle$ is of order
\begin{equation}
    \frac{1}{R^{\Delta_a+\Delta_b}}
    \left(\frac{r}{R}\right)^{-\Delta_c-\Delta_d}.
    \label{eq:first_contribution_in_projected_covariance}
\end{equation}
Second, if $\Delta_c<\Delta_a$ and $\Delta_d\geq\Delta_b$, the corresponding contribution is of order
\begin{equation}
    \frac{1}{R^{\Delta_a+\Delta_d}}
    \left(\frac{r}{R}\right)^{-\Delta_c-\Delta_d}.
    \label{eq:second_contribution_in_projected_covariance}
\end{equation}
The same statement, with the two sides interchanged, applies when $\Delta_c\geq\Delta_a$ and $\Delta_d<\Delta_b$. Third, if $\Delta_c\geq\Delta_a$ and $\Delta_d\geq\Delta_b$, the contribution is of order
\begin{equation}
    \frac{1}{R^{\Delta_c+\Delta_d}}
    \left(\frac{r}{R}\right)^{-\Delta_c-\Delta_d}.
    \label{eq:third_contribution_in_projected_covariance}
\end{equation}
The forms of Eqs.~\eqref{eq:first_contribution_in_projected_covariance}--\eqref{eq:third_contribution_in_projected_covariance} naturally suggest a scaling form in Eq.~\eqref{eq:scaling_form_of_projected_covariance}. Given a fixed $r/R$ and large enough $R$, all three classes of contributions are bounded by $R^{-\Delta_a-\Delta_b}$, multiplied by powers of $r/R$. We therefore obtain the scaling form
\begin{equation}
    \mathcal C_{ab}(r;R) = R^{-\Delta_a-\Delta_b}\left[\widetilde\Phi_{ab}\left(\frac{r}{R}\right)+\ldots\right],
    \qquad R\rightarrow+\infty.
\end{equation}
where $\ldots$ vanishes in the large-$R$ limit. Consequently, this gives $\rho_{ab}$ in Eq.~\eqref{eq:scaling_form_of_projected_covariance}, with
\begin{equation}
    \Phi_{ab}(r/R)=\frac{\widetilde\Phi_{ab}(r/R)}{\sqrt{|\widetilde\Phi_{aa}(r/R)\widetilde\Phi_{bb}(r/R)|}}.
\end{equation}

At last, we briefly discuss what happens to $\rho_{ab}(r;R)$ when at least one of the scaling dimensions $\Delta_a$ and $\Delta_b$ is degenerate. In this situation, diagonalization selects combinations within the corresponding subspace spanned by the scaling operators with the same scaling dimensions. In other words, $\rho_{ab}(r;R)$ no longer compares two uniquely-specified scaling operators.

\subsection{Cross Correlations Between Conformal Towers}
\label{app:cross_tower_collapse}

The derivation in Appendix~\ref{app:eigenoperator_asymptotics} required only the scaling form of the covariance matrix. We now introduce the additional structure of a CFT, in which the scaling operators are organized into conformal towers and plane two-point functions between distinct towers vanish. We consider two nondegenerate branches $a$ and $b$, as in Eq.~\eqref{eq:rho_scaling_limit}. For each branch, we separate the scaling fields below its target dimension from those at or above it. Substituting Eq.~\eqref{eq:eigenoperator_in_scaling_operator_expression} into Eq.~\eqref{eq:projected_covariance} at $r/R=\mathrm{const.}$ gives the exact decomposition

\begin{widetext}
\begin{equation}
\begin{aligned}
\mathcal C_{ab}(r;R)
=R^{-\Delta_a-\Delta_b}\Bigg[{}&
\sum_{\substack{\Delta_c<\Delta_a\\ \Delta_d<\Delta_b}}
A_{ca}^{*}(R)A_{db}(R)K_{cd}(r)\left(\frac{r}{R}\right)^{-\Delta_c-\Delta_d}
\\
&+\sum_{\substack{\Delta_c<\Delta_a\\ \Delta_d\geq\Delta_b}}
R^{-(\Delta_d-\Delta_b)}A_{ca}^{*}(R)q_{db}(R)K_{cd}(r)\left(\frac{r}{R}\right)^{-\Delta_c-\Delta_d}
\\
&+\sum_{\substack{\Delta_c\geq\Delta_a\\ \Delta_d<\Delta_b}}
R^{-(\Delta_c-\Delta_a)}q_{ca}^{*}(R)A_{db}(R)K_{cd}(r)\left(\frac{r}{R}\right)^{-\Delta_c-\Delta_d}
\\
&+\sum_{\substack{\Delta_c\geq\Delta_a\\ \Delta_d\geq\Delta_b}}
R^{-(\Delta_c-\Delta_a)-(\Delta_d-\Delta_b)}q_{ca}^{*}(R)q_{db}(R)K_{cd}(r)\left(\frac{r}{R}\right)^{-\Delta_c-\Delta_d}
\Bigg].
\end{aligned}
\label{eq:cross_tower_full_expansion_app}
\end{equation}
\end{widetext}

The four sums distinguish whether the scaling dimensions on the two sides lie below or at and above their respective target dimensions. Their explicit powers of $R$ identify which terms can survive at the leading order $R^{-\Delta_a-\Delta_b}$. To collect the coefficients of these terms, define
\begin{equation}
    t_{ca}=
    \begin{cases}
        \displaystyle\lim_{R\rightarrow\infty}A_{ca}(R),
        & \Delta_c<\Delta_a,\\[4pt]
        \displaystyle\lim_{R\rightarrow\infty}q_{aa}(R),
        & c=a.
    \end{cases}
    \label{eq:leading_eigenoperator_coefficients_app}
\end{equation}
The leading function of the unnormalized correlation can then be written as
\begin{equation}
    \widetilde\Phi_{ab}(r/R)
    =\sum_{\substack{\Delta_c\leq\Delta_a\\ \Delta_d\leq\Delta_b}}
    t_{ca}^{*}t_{db}k_{cd}\left(\frac{r}{R}\right)^{-\Delta_c-\Delta_d}.
    \label{eq:leading_cross_correlation_function_app}
\end{equation}

We next determine which scaling fields can have nonzero coefficients $t_{ca}$. For a nondegenerate branch $a$, Eq.~\eqref{eq:eigenoperator_lower_limit_app} reduces to
\begin{equation}
    t_{<a,a}=-k_{<a,<a}^{-1}k_{<a,a}t_{aa},
    \label{eq:cross_tower_lower_coefficients_app}
\end{equation}
where $<a$ denotes the scaling fields with dimensions below $\Delta_a$. Suppose that $\phi_a$ belongs to a conformal tower $\mathcal T_a$. Because plane two-point functions between distinct conformal towers vanish, $k_{<a,<a}$ is block diagonal in the tower decomposition, and its inverse preserves the same blocks. Moreover, $k_{<a,a}$ can be nonzero only for fields in $\mathcal T_a$. Equation~\eqref{eq:cross_tower_lower_coefficients_app} then implies
\begin{equation}
    t_{ca}=0
    \qquad\text{if}\qquad
    \phi_c\notin\mathcal T_a.
    \label{eq:leading_coefficient_tower_selection_app}
\end{equation}
Thus, although the orthogonal lattice operator contains many lower-dimensional scaling fields, its leading coefficients can be nonzero only for fields in the conformal tower of its target field.

\begin{table*}[t]
\caption{Point sets used to construct the even-sector operators. The last column lists the lattice spins $\tilde{s}$ for which the operator $J_{\tilde{s}}E(\mathbf x;X_{\mu})$ is nonzero.}
\label{tab:square_energy_lattice_operators}
\begin{ruledtabular}
\begin{tabular}{cll}
 & points in $X_{\mu}$ & allowed lattice spins $\tilde{s}$ \\
\hline
$X_1$ & $\{\mathbf0,\mathbf e_1\}$ & $0,1,2,3$ \\
$X_2$ & $\{\mathbf0,\mathbf e_1+\mathbf e_2\}$ & $0,1,2,3$ \\
$X_3$ & $\{-\mathbf e_1,\mathbf e_1\}$ & $0,2$ \\
$X_4$ & $\{-\mathbf e_1-\mathbf e_2,\mathbf e_1+\mathbf e_2\}$ & $0,2$ \\
$X_5$ & $\{\mathbf0,\mathbf e_1,\mathbf e_2,\mathbf e_1+\mathbf e_2\}$ & $0,1,2,3$ \\
$X_6$ & $\{-\mathbf e_1,-\mathbf e_2,\mathbf e_1,\mathbf e_2\}$ & $0$ \\
$X_7$ & $\{-\mathbf e_1-\mathbf e_2,-\mathbf e_1+\mathbf e_2,\mathbf e_1-\mathbf e_2,\mathbf e_1+\mathbf e_2\}$ & $0$ \\
\end{tabular}
\end{ruledtabular}
\end{table*}

Now let $\phi_a$ and $\phi_b$ belong to distinct conformal towers $\mathcal T_a$ and $\mathcal T_b$. Every potentially nonzero term in Eq.~\eqref{eq:leading_cross_correlation_function_app} has $\phi_c\in\mathcal T_a$ and $\phi_d\in\mathcal T_b$. The two-point amplitude $k_{cd}$ therefore vanishes, and it follows that
\begin{equation}
    \widetilde\Phi_{ab}(r/R)=0
    \label{eq:cross_tower_leading_function_vanishes_app}
\end{equation}
for every fixed $r/R$, and hence $\Phi_{ab}(r/R)=0$ as well. This conclusion is stronger than the orthogonality at $r/R=1$ imposed by diagonalizing the covariance matrix: the complete leading scaling function vanishes because the two branches approach distinct conformal towers.

The cross correlation is therefore governed by a subleading contribution. Let $g>0$ denote the smallest additional power of $R^{-1}$ with a nonzero coefficient for the pair under consideration. This power may arise from a higher-dimensional scaling field, a finite-$R$ correction to the eigenvector coefficients, or the irrelevant correction in $K_{cd}(R)$. We then have
\begin{equation}
    \mathcal C_{ab}(r;R)
    =R^{-\Delta_a-\Delta_b-g}
    \bigl[\widetilde F^{(1)}_{ab}(r/R)+\ldots\bigr].
    \label{eq:cross_tower_subleading_scaling_app}
\end{equation}
In contrast, the diagonal correlations retain their leading scaling forms,
\begin{equation}
    \begin{aligned}
        \mathcal C_{aa}(r;R)
        &=R^{-2\Delta_a}\bigl[\widetilde\Phi_{aa}(r/R)+\ldots\bigr],\\
        \mathcal C_{bb}(r;R)
        &=R^{-2\Delta_b}\bigl[\widetilde\Phi_{bb}(r/R)+\ldots\bigr].
    \end{aligned}
    \label{eq:cross_tower_diagonal_scaling_app}
\end{equation}
Substitution into Eq.~\eqref{eq:rho_definition} gives
\begin{equation}
    \lim_{R\rightarrow\infty}R^g\rho_{ab}(r;R)
    =\frac{\widetilde F^{(1)}_{ab}(r/R)}
    {\sqrt{\left|\widetilde\Phi_{aa}(r/R)\widetilde\Phi_{bb}(r/R)\right|}}.
    \label{eq:cross_tower_rescaled_collapse_app}
\end{equation}
The ratio on the right-hand side defines the normalized correction function $F^{(1)}_{ab}(r/R)$ used in the numerical analysis. Therefore, the unrescaled cross-tower correlation approaches zero and need not exhibit a nontrivial collapse. The rescaled quantity $R^g\rho_{ab}(r;R)$ can instead collapse to a nonzero function of $r/R$ determined by the first surviving correction.

\section{Square Lattice Results of the Operator Covariance Method}
\label{app:square}

We now apply the same analysis to the infinite square lattice, whose critical coupling is $K_c=\frac12\ln(1+\sqrt2)$. Since the rotational symmetry of the square lattice is $C_4$, we introduce the four-fold $(\pi/2)$-rotation $Q_4$ on the plane, and it acts on a point set $X$ by
\begin{equation}
    Q_4^q[X]=\{Q_4^q\boldsymbol\delta:\boldsymbol\delta\in X\}.
    \label{eq:square_rotated_point_set}
\end{equation}
Similar to what we have implemented on the triangular lattice, here we also construct the lattice operators with definite lattice spin $\tilde{s}$:
\begin{equation}
    J_{\tilde{s}}E(\mathbf x;X)=\frac{\sqrt{\ell_X}}{4}\sum_{q=0}^{3}
    e^{-2\pi i \tilde{s} q/4}E(\mathbf x;Q_4^q[X]),
    \label{eq:square_lattice_spin_operator}
\end{equation}
where $\tilde{s}\in\mathbb Z/4\mathbb Z$ denotes lattice spin and $\ell_X$ now counts the distinct point sets generated by $Q_4$, with the same normalization over distinct orientations as on the triangular lattice. For the square-lattice bases used here, Eqs.~\eqref{eq:bilinear_correlation_of_lattice_operators}--\eqref{eq:triangular_complex_covariance}, with $P_{\tilde{s}}$ replaced by $J_{\tilde{s}}$, define a Hermitian covariance matrix. We measure correlations along a lattice axis and fit the scaling dimensions over $96\leq r\leq192$.

The main change in interpreting the spectrum is that continuum spin is now resolved modulo four. For example, fields with continuum spins $s=4$ and $s=-4$ both belong to the scalar lattice sector $\tilde{s}=0$, whereas they belong to $\tilde{s}=4$ and $\tilde{s}=2$ on the triangular lattice. The same continuum CFT can therefore produce different groups of degenerate eigenvalue branches when classified by lattice spin. The selected lattice operators also differ between the two geometries, so the number of resolved directions depends on their overlaps as well as on the rotational symmetry.

\begin{figure*}[t]
    \centering
    \includegraphics[width=\textwidth]{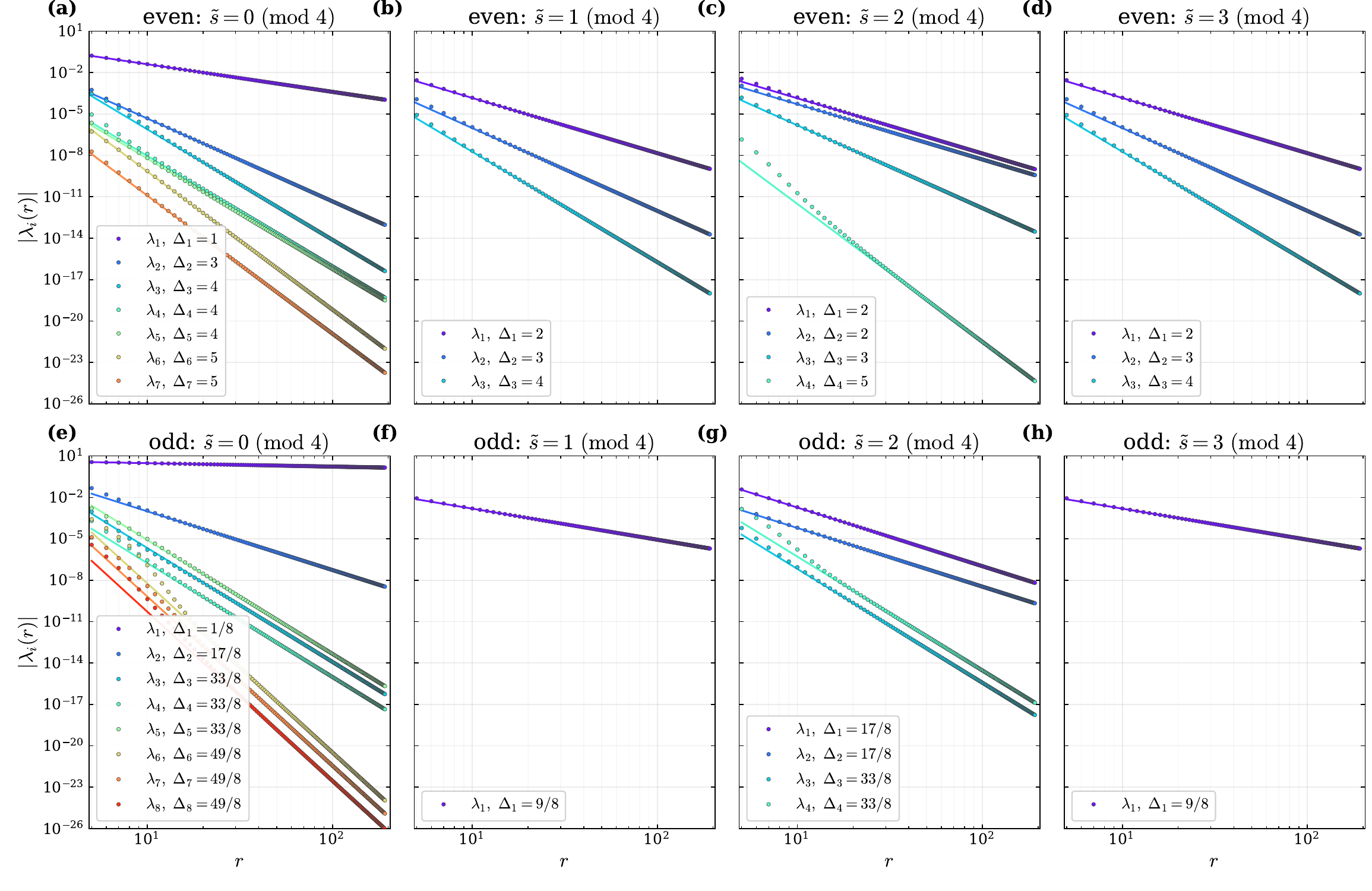}
    \caption{Covariance spectra on the infinite square lattice. Panels (a)--(d) show the even sectors with $\tilde{s}=0,1,2,3\pmod{4}$ for $3\leq r\leq192$, and panels (e)--(h) show the corresponding odd sectors for $5\leq r\leq192$. The dots are the numerically exact plane data and are displayed until a branch first falls below $10^{-26}$. The solid line associated with each branch has the theoretical slope $-2\Delta$ and passes through its last displayed point; these lines are not fits. The legends identify the eigenvalue branches and their theoretical dimensions. The fitted scaling dimensions from the data over $96\leq r\leq192$ are listed in Tables~\ref{tab:square_energy_c4} and~\ref{tab:square_spin_c4} for the even and odd sectors, respectively.}
    \label{fig:square_c4_spectra}
\end{figure*}

\begin{table*}[t]
\caption{Even-sector CFT states associated with the resolved eigenvalue branches. The lattice spin $\tilde{s}$ is understood modulo four. Fitted dimensions are obtained from the data over $96\leq r\leq192$ in Fig.~\ref{fig:square_c4_spectra}(a)--(d). Fitting errors (as parenthesized) are formal regression standard errors.}
\label{tab:square_energy_c4}
\begin{ruledtabular}
\begin{tabular}{ccll}
$\tilde{s}\ (\mathrm{mod}\ 4)$ & $\Delta$ & CFT states & $\Delta_{\rm fit}$ \\
\hline
$0$ & 1 & $\varepsilon$ & $1.0000660(12)$ \\
$0$ & 3 & $L_{-1}\bar L_{-1}\varepsilon$ & $3.0002244(42)$ \\
$0$ & 4 & $L_{-4}\mathbb I,\ L_{-2}^{2}\mathbb I,\ L_{-2}\bar L_{-2}\mathbb I,\ \bar L_{-4}\mathbb I,\ \bar L_{-2}^{2}\mathbb I$ & $4.001530(29),\ 4.001698(32),\ 4.0005271(99)$ \\
$0$ & 5 & $L_{-4}\varepsilon,\ L_{-3}L_{-1}\varepsilon,\ L_{-2}\bar L_{-2}\varepsilon,\ \bar L_{-4}\varepsilon,\ \bar L_{-3}\bar L_{-1}\varepsilon$ & $5.000678(13),\ 5.000682(13)$ \\
$1$ & 2 & $L_{-1}\varepsilon$ & $2.0002322(43)$ \\
$1$ & 3 & $\bar L_{-3}\mathbb I$ & $3.000803(15)$ \\
$1$ & 4 & $L_{-2}\bar L_{-1}\varepsilon,\ \bar L_{-3}\varepsilon$ & $4.000787(15)$ \\
$2$ & 2 & $L_{-2}\mathbb I,\ \bar L_{-2}\mathbb I$ & $2.000831(16),\ 2.0003872(72)$ \\
$2$ & 3 & $L_{-2}\varepsilon,\ \bar L_{-2}\varepsilon$ & $3.0002911(55)$ \\
$2$ & 5 & $L_{-3}\bar L_{-1}\varepsilon,\ L_{-1}\bar L_{-3}\varepsilon$ & $5.01342(25)$ \\
$3$ & 2 & $\bar L_{-1}\varepsilon$ & $2.0002322(43)$ \\
$3$ & 3 & $L_{-3}\mathbb I$ & $3.000803(15)$ \\
$3$ & 4 & $L_{-3}\varepsilon,\ L_{-1}\bar L_{-2}\varepsilon$ & $4.000787(15)$ \\
\end{tabular}
\end{ruledtabular}
\end{table*}

\begin{table*}[t]
\caption{Point sets used to construct the odd-sector operators. The last column lists the lattice spins $\tilde{s}$ for which the operator $J_{\tilde{s}}E(\mathbf x;X_{\mu})$ is nonzero.}
\label{tab:square_spin_lattice_operators}
\begin{ruledtabular}
\begin{tabular}{cll}
 & points in $X_{\mu}$ & allowed lattice spins $\tilde{s}$ \\
\hline
$X_1$ & $\{\mathbf0\}$ & $0$ \\
$X_2$ & $\{-\mathbf e_1,\mathbf0,\mathbf e_1\}$ & $0,2$ \\
$X_3$ & $\{-\mathbf e_1-\mathbf e_2,\mathbf0,\mathbf e_1+\mathbf e_2\}$ & $0,2$ \\
$X_4$ & $\{\mathbf0,-\mathbf e_1-\mathbf e_2,-\mathbf e_1+\mathbf e_2,\mathbf e_1-\mathbf e_2,\mathbf e_1+\mathbf e_2\}$ & $0$ \\
$X_5$ & $\{\mathbf e_1-\mathbf e_2,\mathbf e_1,\mathbf e_1+\mathbf e_2\}$ & $0,1,2,3$ \\
$X_6$ & $\{-2\mathbf e_1-2\mathbf e_2,\mathbf0,2\mathbf e_1+2\mathbf e_2\}$ & $0,2$ \\
$X_7$ & $\{\mathbf0,-2\mathbf e_1-2\mathbf e_2,-2\mathbf e_1+2\mathbf e_2,2\mathbf e_1-2\mathbf e_2,2\mathbf e_1+2\mathbf e_2\}$ & $0$ \\
$X_8$ & $\{\mathbf0,\pm\mathbf e_1\pm\mathbf e_2,\pm2\mathbf e_1\pm2\mathbf e_2\}$ & $0$ \\
\end{tabular}
\end{ruledtabular}
\end{table*}

\subsubsection{Even sector}
\label{sec:ising_square_energy_sector}

In the even sector, we use the seven point sets $X_1,\ldots,X_7$ listed in Table~\ref{tab:square_energy_lattice_operators} to construct the corresponding operators $J_{\tilde{s}}E(\mathbf x;X_{\mu})$. Here $\mathbf e_1=(1,0)$ and $\mathbf e_2=(0,1)$ denote the two primitive lattice directions. The four lattice-spin sectors have dimensions $(7,3,5,3)$ for $\tilde{s}=0,1,2,3$, respectively, and account for 18 lattice operators.

The first row of Fig.~\ref{fig:square_c4_spectra} shows the resulting eigenvalue spectra. As a representative check on how to understand the resolved spectra in general, we examine the $\tilde{s}=0$ sector, whose seven branches approach dimensions $1$, $3$, three copies of $4$, and two copies of $5$. The first two correspond to the energy primary $\varepsilon$ and its Laplacian descendant $L_{-1}\bar L_{-1}\varepsilon$. At $\Delta=4$, the allowed CFT subspace contains $T\bar T$ together with the spin-$\pm4$ descendants of the identity. The three observed branches therefore resolve three independent directions in this five-dimensional subspace. The two remaining branches approach $\Delta=5$ and belong to the energy tower.

The sectors with nonzero lattice spin contain further descendants of $\varepsilon$ and $\mathbb I$. The sectors $\tilde{s}=1$ and $\tilde{s}=3$ each resolve dimensions $2$, $3$, and $4$. Reflection relates these two sectors and gives their coincident spectra.

The $\tilde{s}=2$ sector contains five lattice operators, but interestingly, only four nonzero eigenvalue branches at non-overlapping separations. Its two leading branches approach $\Delta=2$ and are consistent with the stress-tensor components $T=L_{-2}\mathbb I$ and $\bar T=\bar L_{-2}\mathbb I$. The remaining two branches approach dimensions $3$ and $5$. The missing branch results from an exact local identity of the square-lattice Ising model. For any observable $\mathcal A$ that does not contain the spin at $\mathbf{x}$, integrating out this spin $S(\mathbf{x})$ gives
\begin{equation}
    \left\langle J_2E(\mathbf x;X_1)\mathcal A\right\rangle
    =\frac{\tanh(2K_c)}{\sqrt{2}}
    \left\langle J_2E(\mathbf x;X_3)\mathcal A\right\rangle.
    \label{eq:square_ising_null_identity}
\end{equation}
Since $\tanh(2K_c)=1/\sqrt{2}$, the combination
\begin{equation}
    J_2E(\mathbf x;X_1)
    -\frac{1}{2}J_2E(\mathbf x;X_3)
    \label{eq:square_ising_null_operator}
\end{equation}
has vanishing correlations with every non-overlapping lattice operator and therefore forms an exact null direction of the covariance matrix. However, this null direction is accidental in the sense that the exact identity is remarkably special. It causes the only one missing branch in the $\tilde{s}=2$ sector, and no analogous identity and missing branch can be identified in the rest of analysis on both the triangular and the square lattice.

\begin{table*}[t]
\caption{Odd-sector CFT states associated with the resolved eigenvalue branches. The lattice spin $\tilde{s}$ is understood modulo four. Fitted dimensions are obtained from the data over $96\leq r\leq192$ in Fig.~\ref{fig:square_c4_spectra}(e)--(h). Fitting errors (as parenthesized) are formal regression standard errors.}
\label{tab:square_spin_c4}
\footnotesize
\begin{ruledtabular}
\begin{tabular}{ccll}
$\tilde{s}\ (\mathrm{mod}\ 4)$ & $\Delta$ & CFT states & $\Delta_{\rm fit}$ \\
\hline
$0$ & $1/8$ & $\sigma$ & $0.12501711(32)$ \\
$0$ & $17/8$ & $L_{-1}\bar L_{-1}\sigma$ & $2.1251781(35)$ \\
$0$ & $33/8$ & $L_{-4}\sigma,\ L_{-3}L_{-1}\sigma,\ L_{-2}\bar L_{-2}\sigma,\ \bar L_{-4}\sigma,\ \bar L_{-3}\bar L_{-1}\sigma$ & $4.123359(31),\ 4.127710(51),\ 4.127521(47)$ \\
$0$ & $49/8$ & $L_{-5}\bar L_{-1}\sigma,\ L_{-4}L_{-1}\bar L_{-1}\sigma,\ L_{-3}L_{-2}\bar L_{-1}\sigma,\ L_{-3}\bar L_{-3}\sigma,\ L_{-3}\bar L_{-2}\bar L_{-1}\sigma,$ & $6.127808(56),\ 6.129284(82),\ 6.126561(45)$ \\
 & & $L_{-2}L_{-1}\bar L_{-3}\sigma,\ L_{-2}L_{-1}\bar L_{-2}\bar L_{-1}\sigma,\ L_{-1}\bar L_{-5}\sigma,\ L_{-1}\bar L_{-4}\bar L_{-1}\sigma,\ L_{-1}\bar L_{-3}\bar L_{-2}\sigma$ & \\
$1$ & $9/8$ & $L_{-1}\sigma$ & $1.1251301(24)$ \\
$2$ & $17/8$ & $L_{-2}\sigma,\ \bar L_{-2}\sigma$ & $2.1255299(99),\ 2.1252257(42)$ \\
$2$ & $33/8$ & $L_{-3}\bar L_{-1}\sigma,\ L_{-2}L_{-1}\bar L_{-1}\sigma,\ L_{-1}\bar L_{-3}\sigma,\ L_{-1}\bar L_{-2}\bar L_{-1}\sigma$ & $4.126339(25),\ 4.1250147(18)$ \\
$3$ & $9/8$ & $\bar L_{-1}\sigma$ & $1.1251301(24)$ \\
\end{tabular}
\end{ruledtabular}
\end{table*}

After elaborating the special case of the missing branch in the $\tilde{s}=2$ sector, we now turn to the general discussion of the number of the resolved CFT states. Most of the resolved multiplicities follow directly from the numbers of available CFT states and lattice operators. At a given scaling dimension, the eigenvalue branches either exhaust the corresponding CFT subspace or, at the largest resolved dimension in a lattice-spin sector, are limited by the remaining number of independent lattice operators. There are, however, exceptions in the $\tilde{s}=0$ and $\tilde{s}=2$ sectors which are similar to the case discussed in Sec.~\ref{sec:ising_spin_sector}. The incomplete multiplicities in these two can be similarly understood by further inspecting reflection symmetry. Reflection through the measurement axis exchanges $L_{-n}$ and $\bar L_{-n}$ while sending both lattice spins $\tilde{s}=0$ and $\tilde{s}=2$ to themselves. The lattice operators and CFT states in these sectors can therefore be reflection-even or reflection-odd. All seven lattice operators in the $\tilde{s}=0$ sector are reflection-even. At $\Delta=4$, the three reflection-even CFT states are
\begin{equation}
L_{-2}\bar L_{-2}\mathbb I,\qquad
(L_{-4}+\bar L_{-4})\mathbb I,\qquad
(L_{-2}^{2}+\bar L_{-2}^{2})\mathbb I,
\end{equation}
whereas $(L_{-4}-\bar L_{-4})\mathbb I$ and $(L_{-2}^{2}-\bar L_{-2}^{2})\mathbb I$ are reflection-odd. The scalar lattice operators can therefore overlap with only the three even combinations, in agreement with the three observed branches.

This analysis procedure of the $\tilde{s}=2$ sector is similar. Here we only consider the operators constructed from $X_1$, $X_2$, $X_4$, and $X_5$ given the null direction involving $X_1$ and $X_3$. The operators constructed from $X_1$ is reflection-even, whereas those constructed from $X_2$, $X_4$, and $X_5$ are reflection-odd. At $\Delta=2$, the stress-tensor combinations $(L_{-2}+\bar L_{-2})\mathbb I$ and $(L_{-2}-\bar L_{-2})\mathbb I$ are respectively reflection-even and reflection-odd, so the leading pair uses one direction of each type. The remaining two odd directions reveal $(L_{-2}-\bar L_{-2})\varepsilon$ at $\Delta=3$ and $(L_{-3}\bar L_{-1}-L_{-1}\bar L_{-3})\varepsilon$ at $\Delta=5$. The observed multiplicities therefore follow from the reflection properties of the chosen lattice operators.

Table~\ref{tab:square_energy_c4} lists the CFT states and fitted dimensions for these branches. Each fitted value represents a resolved eigenvalue branch, while the states in the same row span the corresponding CFT subspace. Similar to what has already been seen in the case with the triangular lattice, although the fitted dimensions are close to their theoretical values, the discrepancies are much larger than the formal fitting errors, indicating systematic corrections from operator mixing and irrelevant perturbations.

\subsubsection{Odd sector}
\label{sec:ising_square_spin_sector}

For the odd sector, we use the eight point sets $X_1,\ldots,X_8$ listed in Table~\ref{tab:square_spin_lattice_operators} to construct the corresponding operators $J_{\tilde{s}}E(\mathbf x;X_{\mu})$. The four lattice-spin sectors have dimensions $(8,1,4,1)$ for $\tilde{s}=0,1,2,3$, respectively, and account for 14 lattice operators.

The second row of Fig.~\ref{fig:square_c4_spectra} shows the resulting eigenvalue spectra. As a representative check, here we also start with examining the $\tilde{s}=0$ sector, whose eight branches approach dimensions $1/8$, $17/8$, three copies of $33/8$, and three copies of $49/8$. The first two correspond to the spin primary $\sigma$ and its Laplacian descendant $L_{-1}\bar L_{-1}\sigma$. At $\Delta=33/8$, the allowed CFT subspace contains states with continuum spins $s=0$ and $s=\pm4$, which all belong to the same lattice-spin sector modulo four. The three observed branches therefore resolve three independent directions in this five-dimensional subspace spanned by the CFT states.

The sectors with nonzero lattice spin contain further descendants of $\sigma$. The $\tilde{s}=2$ sector resolves two branches at $\Delta=17/8$ and two at $\Delta=33/8$. Its leading pair is consistent with the level-two descendants of $\sigma$ with continuum spins $s=2$ and $s=-2$, which belong to the same lattice-spin sector modulo four. The $\tilde{s}=1$ and $\tilde{s}=3$ sectors each resolve one branch at $\Delta=9/8$, corresponding to the two first derivatives of the spin field. Reflection relates these two sectors and gives their coincident spectra.

\begin{figure*}[t]
    \centering
    \includegraphics[width=\textwidth]{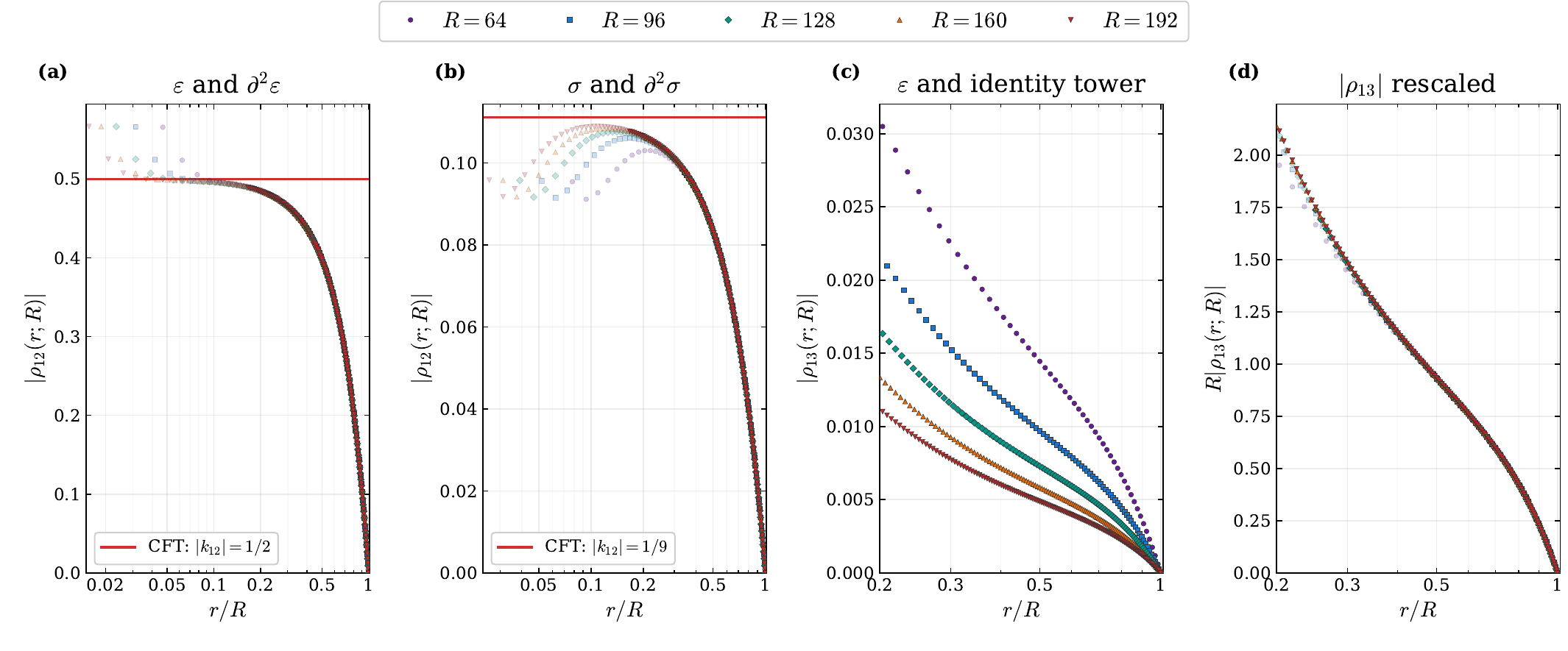}
    \caption{Finite-distance tests of the covariance eigenvectors in the $\tilde{s}=0$ sectors on the infinite square lattice, plotted against $r/R$ for $R=64,96,128,160,$ and $192$. Panels (a) and (b) show $|\rho_{12}(r;R)|$ for $(\varepsilon,\partial^2\varepsilon)$ in the even sector and $(\sigma,\partial^2\sigma)$ in the odd sector, respectively. The horizontal red lines mark the plane-CFT two-point amplitudes $|k_{12}|=1/2$ and $1/9$. Panels (c) and (d) show the cross-tower correlation between $\varepsilon$ and a dimension-four state in the identity tower: panel (c) displays the unrescaled magnitude $|\rho_{13}(r;R)|$, while panel (d) displays $R|\rho_{13}(r;R)|$. These two panels use $r/R\geq0.2$. Every curve vanishes at $r/R=1$ by construction. Points with $r<32$ are shaded in panels (a), (b), and (d) to indicate the short-distance region but are not excluded; the unrescaled data in panel (c) are shown without shading.}
    \label{fig:ising_square_eigenvector_test}
\end{figure*}

Similar to all the previous examples, most of the resolved multiplicities here follow directly from the numbers of available CFT states and lattice operators. The exceptions in the $\tilde{s}=0$ and $\tilde{s}=2$ sectors can again be understood by further inspecting reflection symmetry. All eight lattice operators in the $\tilde{s}=0$ sector are reflection-even. At $\Delta=33/8$, the three reflection-even CFT states are
\begin{equation}
L_{-2}\bar L_{-2}\sigma,\qquad
(L_{-4}+\bar L_{-4})\sigma,\qquad
(L_{-3}L_{-1}+\bar L_{-3}\bar L_{-1})\sigma,
\end{equation}
whereas the two corresponding differences are reflection-odd. The scalar lattice operators can therefore overlap with only the three even combinations, in agreement with the three observed branches. After the branches at $\Delta=1/8$, $17/8$, and $33/8$ have been resolved, only three lattice directions remain, accounting for the three observed branches at $\Delta=49/8$.

The analysis procedure of the $\tilde{s}=2$ sector is similar. The operators constructed from $X_2$ and $X_5$ are reflection-even, while those constructed from $X_3$ and $X_6$ are reflection-odd. At $\Delta=17/8$, the combinations $(L_{-2}+\bar L_{-2})\sigma$ and $(L_{-2}-\bar L_{-2})\sigma$ are respectively reflection-even and reflection-odd, so the leading pair uses one direction of each type. The two remaining directions, one even and one odd, then resolve two of the four CFT states at $\Delta=33/8$. This completes the analysis on the incomplete multiplicities between the CFT states and the number of the resolved scaling dimensions.

Table~\ref{tab:square_spin_c4} lists the CFT states after removal of null states and gives one fitted value for each resolved branch.

\subsubsection{Extraction of $k_{ab}$ for the Square Lattice}
\label{sec:ising_square_eigenvector_test}

After analyzing the eigenvalue spectra, we now demonstrate how the two-point amplitudes from the covariance eigenvectors can be extracted. We restrict this test to the lattice-spin $\tilde{s}=0$ sectors in the even and odd sectors. Similar to what has been implemented in the case of the triangular lattice, at each construction distance $R$, we diagonalize the covariance matrix, use its eigenvectors to form the orthogonal lattice operators in Eq.~\eqref{eq:eigenoperator}, and measure their normalized cross ratios $\rho_{ab}(r;R)$ in Eq.~\eqref{eq:rho_definition} at separations $r$.

We first study the magnitude of the normalized cross ratio $|\rho_{12}(r;R)|$ between the leading two scalar branches. In the lattice-spin $\tilde{s}=0$ sectors, these are $(\varepsilon,\partial^2\varepsilon)$ in the even sector and $(\sigma,\partial^2\sigma)$ in the odd sector. Equation~\eqref{eq:ising_rho_cft} therefore gives the predicted amplitudes $|k_{12}|=1/2$ and $|k_{12}|=1/9$, respectively.

To demonstrate how $|k_{12}|$ can be extracted from the normalized cross ratio $|\rho_{12}(r;R)|$, we perform a scaling-collapse analysis by plotting $|\rho_{12}(r;R)|$ against $r/R$. Panels (a) and (b) of Fig.~\ref{fig:ising_square_eigenvector_test} show the results for $R=64,96,128,160,$ and $192$. In both the even and odd sectors, the curves approach a common profile over intermediate $r/R$ and agree well with the corresponding CFT amplitudes. At $R=192$, agreement within five percent is obtained over $0.021\lesssim r/R\leq0.25$ in the even sector and $0.063\lesssim r/R\lesssim0.22$ in the odd sector. Points with $r<32$ are shaded to distinguish the short-distance region but remain in the plot. The alignment of the curves becomes clearer as $R$ increases, while their common zero at $r/R=1$ follows from the orthogonality imposed by diagonalization. Importantly, with the scaling collapse analysis, we demonstrate that the values $|k_{12}|$ for both the even and the odd sectors are invariant under different lattice geometry. Hence, this indicates the universal property of the two-point amplitudes.

As the next nontrivial test of the prediction in Eq.~\eqref{eq:rho_scaling_limit}, we study the orthogonality of fields from different conformal towers. In the $\tilde{s}=0$ sector of the even sector, the first branch approaches $\varepsilon$, while the third approaches a dimension-four state in the identity tower. Since these fields belong to different conformal towers, their plane two-point amplitude vanishes: $k_{13}=0$ \cite{DiFrancesco1997}. Through measuring $|\rho_{13}(r;R)|$, we can therefore examine the orthogonality between these fields.

We first plot $|\rho_{13}(r;R)|$ against $r/R$ in panel (c) of Fig.~\ref{fig:ising_square_eigenvector_test}. Since the leading continuum contribution to $\mathcal C_{13}(r;R)$ vanishes, the curves do not collapse for different choices of $R$. The next scalar field in the $\varepsilon$ tower above dimension four is $L_{-2}\bar L_{-2}\varepsilon$ at $\Delta=5$, which can correlate with $\varepsilon$. As result, the first allowed correction has the form $|\rho_{13}(r;R)|=R^{-1}|F^{(1)}_{13}(r/R)|+\ldots$. Consequently, the curves of $R|\rho_{13}(r;R)|$ in panel (d) display a good approximate collapse. This behavior is consistent with a vanishing continuum two-point amplitude and a nonzero $R^{-1}$ finite-distance correction.

\bibliography{references}

@article{Luscher1990,
  author  = {L{\"u}scher, Martin and Wolff, Ulli},
  title   = {How to Calculate the Elastic Scattering Matrix
             in Two-Dimensional Quantum Field Theories
             by Numerical Simulation},
  journal = {Nuclear Physics B},
  volume  = {339},
  pages   = {222--252},
  year    = {1990}
}

@book{Cardy1996,
  author    = {Cardy, John},
  title     = {Scaling and Renormalization in Statistical Physics},
  publisher = {Cambridge University Press},
  address   = {Cambridge},
  year      = {1996}
}

@article{Fisher1998,
  author  = {Fisher, Michael E.},
  title   = {Renormalization Group Theory: Its Basis and Formulation in Statistical Physics},
  journal = {Reviews of Modern Physics},
  volume  = {70},
  number  = {2},
  pages   = {653--681},
  year    = {1998},
  doi     = {10.1103/RevModPhys.70.653}
}

@incollection{Wegner1976,
  author    = {Wegner, Franz J.},
  title     = {The Critical State, General Aspects},
  booktitle = {Phase Transitions and Critical Phenomena},
  editor    = {Domb, Cyril and Green, Melville S.},
  volume    = {6},
  pages     = {7--124},
  publisher = {Academic Press},
  address   = {London},
  year      = {1976}
}

@article{Sandvik2026,
  author    = {Sandvik, Anders W.},
  title     = {Using Operator Covariance to Disentangle Scaling Dimensions in Lattice Models},
  journal   = {Physical Review E},
  volume    = {114},
  pages     = {024146},
  year      = {2026},
  publisher = {American Physical Society},
  doi       = {10.1103/jq2v-wjkf}
}

@article{ThomasMiddleton2013,
  author    = {Thomas, Creighton K. and Middleton, A. Alan},
  title     = {Numerically Exact Correlations and Sampling in the Two-Dimensional Ising Spin Glass},
  journal   = {Physical Review E},
  volume    = {87},
  pages     = {043303},
  year      = {2013},
  publisher = {American Physical Society},
  doi       = {10.1103/PhysRevE.87.043303},
  eprint    = {1301.1252},
  archivePrefix = {arXiv}
}

@book{DiFrancesco1997,
  author    = {Di Francesco, Philippe and Mathieu, Pierre and Senechal, David},
  title     = {Conformal Field Theory},
  publisher = {Springer},
  address   = {New York},
  year      = {1997},
  doi       = {10.1007/978-1-4612-2256-9}
}

@misc{Rychkov2016,
  author        = {Rychkov, Slava},
  title         = {{EPFL} Lectures on Conformal Field Theory in {D} $\geq$ 3 Dimensions},
  year          = {2016},
  eprint        = {1601.05000},
  archivePrefix = {arXiv}
}

@misc{SimmonsDuffin2016,
  author        = {Simmons-Duffin, David},
  title         = {{TASI} Lectures on the Conformal Bootstrap},
  year          = {2016},
  eprint        = {1602.07982},
  archivePrefix = {arXiv}
}

@book{BlumenhagenPlauschinn2009,
  author    = {Blumenhagen, Ralph and Plauschinn, Erik},
  title     = {Introduction to Conformal Field Theory: With Applications to String Theory},
  series    = {Lecture Notes in Physics},
  volume    = {779},
  publisher = {Springer},
  address   = {Berlin},
  year      = {2009},
  doi       = {10.1007/978-3-642-00450-6}
}

@book{HornJohnson2013,
  author    = {Horn, Roger A. and Johnson, Charles R.},
  title     = {Matrix Analysis},
  edition   = {2},
  publisher = {Cambridge University Press},
  address   = {Cambridge},
  year      = {2013}
}

@book{Zhang2005,
  author    = {Zhang, Fuzhen},
  title     = {The Schur Complement and Its Applications},
  publisher = {Springer},
  address   = {New York},
  year      = {2005}
}

@article{Stephenson1964,
  author  = {Stephenson, J.},
  title   = {{Ising}-Model Spin Correlations on the Triangular Lattice},
  journal = {Journal of Mathematical Physics},
  volume  = {5},
  number  = {8},
  pages   = {1009--1024},
  year    = {1964},
  doi     = {10.1063/1.1704202}
}

@article{KacWard1952,
  author  = {Kac, Mark and Ward, John C.},
  title   = {A Combinatorial Solution of the Two-Dimensional {Ising} Model},
  journal = {Physical Review},
  volume  = {88},
  number  = {6},
  pages   = {1332--1337},
  year    = {1952},
  doi     = {10.1103/PhysRev.88.1332}
}

@article{Kasteleyn1963,
  author  = {Kasteleyn, P. W.},
  title   = {Dimer Statistics and Phase Transitions},
  journal = {Journal of Mathematical Physics},
  volume  = {4},
  number  = {2},
  pages   = {287--293},
  year    = {1963},
  doi     = {10.1063/1.1703953}
}

@article{Fisher1966,
  author  = {Fisher, Michael E.},
  title   = {On the Dimer Solution of Planar {Ising} Models},
  journal = {Journal of Mathematical Physics},
  volume  = {7},
  number  = {10},
  pages   = {1776--1781},
  year    = {1966},
  doi     = {10.1063/1.1704825}
}

@article{ChelkakCimasoniKassel2017,
  author  = {Chelkak, Dmitry and Cimasoni, David and Kassel, Adrien},
  title   = {Revisiting the Combinatorics of the {2D} {Ising} Model},
  journal = {Annales de l'Institut Henri Poincar{\'e} D},
  volume  = {4},
  number  = {3},
  pages   = {309--385},
  year    = {2017},
  doi     = {10.4171/AIHPD/42}
}

@article{Cserti2000,
  author  = {Cserti, J{\'o}zsef},
  title   = {Application of the Lattice Green's Function for Calculating
             the Resistance of an Infinite Network of Resistors},
  journal = {American Journal of Physics},
  volume  = {68},
  number  = {10},
  pages   = {896--906},
  year    = {2000},
  doi     = {10.1119/1.1285881},
  eprint  = {cond-mat/9909120},
  archivePrefix = {arXiv}
}

@article{Zhugayevych2025,
  author  = {Zhugayevych, Andriy},
  title   = {Efficient Evaluation of Lattice Green's Functions},
  journal = {Journal of Physics A: Mathematical and Theoretical},
  volume  = {58},
  number  = {2},
  pages   = {025209},
  year    = {2025},
  doi     = {10.1088/1751-8121/ada0fb}
}

@article{MontrollPottsWard1963,
  author  = {Montroll, Elliott W. and Potts, Renfrey B. and Ward, John C.},
  title   = {Correlations and Spontaneous Magnetization of the
             Two-Dimensional {Ising} Model},
  journal = {Journal of Mathematical Physics},
  volume  = {4},
  number  = {2},
  pages   = {308--322},
  year    = {1963},
  doi     = {10.1063/1.1703955}
}

@article{ChengWu1967,
  author  = {Cheng, Hung and Wu, Tai Tsun},
  title   = {Theory of {Toeplitz} Determinants and the Spin Correlations
             of the Two-Dimensional {Ising} Model. III},
  journal = {Physical Review},
  volume  = {164},
  number  = {2},
  pages   = {719--735},
  year    = {1967},
  doi     = {10.1103/PhysRev.164.719}
}

@article{Wu1982,
  author  = {Wu, F. Y.},
  title   = {The {Potts} Model},
  journal = {Reviews of Modern Physics},
  volume  = {54},
  pages   = {235--268},
  year    = {1982},
  doi     = {10.1103/RevModPhys.54.235}
}

@article{Wegner1972,
  author  = {Wegner, Franz J.},
  title   = {Corrections to Scaling Laws},
  journal = {Physical Review B},
  volume  = {5},
  number  = {11},
  pages   = {4529--4536},
  year    = {1972},
  doi     = {10.1103/PhysRevB.5.4529}
}

@article{PelissettoVicari2002,
  author  = {Pelissetto, Andrea and Vicari, Ettore},
  title   = {Critical Phenomena and Renormalization-Group Theory},
  journal = {Physics Reports},
  volume  = {368},
  number  = {6},
  pages   = {549--727},
  year    = {2002},
  doi     = {10.1016/S0370-1573(02)00219-3}
}

@article{Cardy1984,
  author  = {Cardy, John L.},
  title   = {Conformal Invariance and Universality in Finite-Size Scaling},
  journal = {Journal of Physics A: Mathematical and General},
  volume  = {17},
  number  = {7},
  pages   = {L385--L387},
  year    = {1984},
  doi     = {10.1088/0305-4470/17/7/003}
}

@article{BelavinPolyakovZamolodchikov1984,
  author  = {Belavin, Alexander A. and Polyakov, Alexander M. and Zamolodchikov, Alexander B.},
  title   = {Infinite Conformal Symmetry in Two-Dimensional Quantum Field Theory},
  journal = {Nuclear Physics B},
  volume  = {241},
  number  = {2},
  pages   = {333--380},
  year    = {1984},
  doi     = {10.1016/0550-3213(84)90052-X}
}

@article{FriedanQiuShenker1984,
  author  = {Friedan, Daniel and Qiu, Zongan and Shenker, Stephen},
  title   = {Conformal Invariance, Unitarity, and Critical Exponents in Two Dimensions},
  journal = {Physical Review Letters},
  volume  = {52},
  number  = {18},
  pages   = {1575--1578},
  year    = {1984},
  doi     = {10.1103/PhysRevLett.52.1575}
}

@incollection{Ginsparg1988,
  author    = {Ginsparg, Paul},
  title     = {Applied Conformal Field Theory},
  booktitle = {Fields, Strings and Critical Phenomena},
  editor    = {Br{\'e}zin, {\'E}douard and Zinn-Justin, Jean},
  series    = {Les Houches Summer School Proceedings},
  volume    = {49},
  pages     = {1--168},
  publisher = {North-Holland},
  address   = {Amsterdam},
  year      = {1990},
  eprint    = {hep-th/9108028},
  archivePrefix = {arXiv}
}

@article{RochaCaridiWallach1984,
  author  = {Rocha-Caridi, Alvany and Wallach, Nolan R.},
  title   = {Characters of Irreducible Representations of the Virasoro Algebra},
  journal = {Mathematische Zeitschrift},
  volume  = {185},
  pages   = {1--21},
  year    = {1984},
  doi     = {10.1007/BF01214971}
}

@article{Onsager1944,
  author  = {Onsager, Lars},
  title   = {Crystal Statistics. {I}. A Two-Dimensional Model with an Order-Disorder Transition},
  journal = {Physical Review},
  volume  = {65},
  number  = {3--4},
  pages   = {117--149},
  year    = {1944},
  doi     = {10.1103/PhysRev.65.117}
}

@book{McCoyWu1973,
  author    = {McCoy, Barry M. and Wu, Tai Tsun},
  title     = {The Two-Dimensional Ising Model},
  publisher = {Harvard University Press},
  address   = {Cambridge, Massachusetts},
  year      = {1973}
}

@article{Kaufman1949,
  author  = {Kaufman, Bruria},
  title   = {Crystal Statistics. {II}. Partition Function Evaluated by Spinor Analysis},
  journal = {Physical Review},
  volume  = {76},
  number  = {8},
  pages   = {1232--1243},
  year    = {1949},
  doi     = {10.1103/PhysRev.76.1232}
}

@article{ChelkakHonglerIzyurov2015,
  author  = {Chelkak, Dmitry and Hongler, Cl{\'e}ment and Izyurov, Konstantin},
  title   = {Conformal Invariance of Spin Correlations in the Planar {Ising} Model},
  journal = {Annals of Mathematics},
  volume  = {181},
  number  = {3},
  pages   = {1087--1138},
  year    = {2015},
  doi     = {10.4007/annals.2015.181.3.5}
}

@inproceedings{Song2022,
  author = {Song, X. Carol and Smith, Preston
            and Kalyanam, Rajesh and Zhu, Xiao
            and Adams, Eric and Colby, Kevin
            and Finnegan, Patrick and Gough, Erik
            and Hillery, Elizabett and Irvine, Rick
            and Maji, Amiya and St. John, Jason},
  title = {{Anvil} -- System Architecture and
           Experiences from Deployment and
           Early User Operations},
  booktitle = {Practice and Experience in Advanced
               Research Computing (PEARC '22)},
  year = {2022},
  publisher = {ACM},
  note = {Article 23},
  pages = {1--9},
  doi = {10.1145/3491418.3530766}
}

@inproceedings{Boerner2023,
  author = {Boerner, Timothy J. and Deems, Stephen
            and Furlani, Thomas R. and Knuth, Shelley L.
            and Towns, John},
  title = {{ACCESS}: Advancing Innovation: {NSF}'s
           Advanced Cyberinfrastructure Coordination
           Ecosystem: Services \& Support},
  booktitle = {Practice and Experience in Advanced
               Research Computing (PEARC '23)},
  pages = {173--176},
  year = {2023},
  publisher = {ACM},
  doi = {10.1145/3569951.3597559}
}

@article{vidal2020,
  title = {Conformal Fields and Operator Product Expansion in Critical Quantum Spin Chains},
  author = {Zou, Yijian and Milsted, Ashley and Vidal, Guifre},
  journal = {Phys. Rev. Lett.},
  volume = {124},
  issue = {4},
  pages = {040604},
  numpages = {5},
  year = {2020},
  month = {Jan},
  publisher = {American Physical Society},
  doi = {10.1103/PhysRevLett.124.040604},
  url = {https://link.aps.org/doi/10.1103/PhysRevLett.124.040604}
}

\end{document}